\documentclass[twocolumn,a4paper,reprint,amssymb,aps,prd,groupedaddress,superscriptaddress,nofootinbib]{revtex4-2}
\pdfoutput=1
\usepackage{graphicx}
\usepackage{epsf}
\usepackage{bm}
\usepackage{amsmath}
\usepackage{amsfonts}
\usepackage{amssymb}
\usepackage{epstopdf}
\usepackage{natbib}
\usepackage{color}
\usepackage[dvipsnames]{xcolor}
\usepackage{physics}
\usepackage[colorlinks = true,
            linkcolor = blue,
            urlcolor  = blue,
            citecolor = blue,
            anchorcolor = blue]{hyperref}
\usepackage{lipsum}
\providecommand{\U}[1]{\protect\rule{.1in}{.1in}}

\usepackage[capitalize]{cleveref}

\usepackage{soul}

\usepackage{braket}

\newcommand{\be}{\begin{equation}}
\newcommand{\ee}{\end{equation}}
\newcommand{\bea}{\begin{eqnarray}}
\newcommand{\eea}{\end{eqnarray}}

\begin{document}

\title{Inevitable manifestation of wiggles in the expansion of the late Universe}

\author{\"{O}zg\"{u}r Akarsu}
\email{akarsuo@itu.edu.tr}
\affiliation{Department of Physics, Istanbul Technical University, Maslak 34469 Istanbul, Turkey}

\author{Eoin \'O Colg\'ain}
\email{eoin.ocolgain@atu.ie}
\affiliation{Atlantic Technological University, Ash Lane, Sligo, Ireland}
\affiliation{Center for Quantum Spacetime, Sogang University, Seoul 121-742, Korea}
\affiliation{Department of Physics, Sogang University, Seoul 121-742, Korea}

\author{Emre \"{O}z\"{u}lker}
\email{ozulker17@itu.edu.tr}
\affiliation{Department of Physics, Istanbul Technical University, Maslak 34469 Istanbul, Turkey}

\author{Somyadip Thakur}
\email{somyadip@sogang.ac.kr}
\affiliation{Center for Quantum Spacetime, Sogang University, Seoul 121-742, Korea}
\affiliation{Department of Physics, Sogang University, Seoul 121-742, Korea}

\author{Lu Yin}
\email{lu.yin@apctp.org}
\affiliation{Center for Quantum Spacetime, Sogang University, Seoul 121-742, Korea}
\affiliation{Department of Physics, Sogang University, Seoul 121-742, Korea}
\affiliation{Asia Pacific Center for Theoretical Physics, Pohang 37673, Korea}

\begin{abstract}
Using the fact that the comoving angular diameter distance to the last scattering surface is strictly constrained almost model independently, we show that, for any model agreeing with the standard $\Lambda$CDM model on its background dynamics at $z\sim0$ and size of the comoving sound horizon at last scattering, the deviations of the Hubble radius from the one of the standard $\Lambda$CDM model must be a member of the set of \textit{admissible wavelets}. The family of models characterized by this framework also offers nontrivial oscillatory behaviours in various functions that define the kinematics of the Universe, even when the wavelets themselves are very simple. We also discuss the consequences of attributing these kinematics to, first, dark energy, and second, varying gravitational coupling strength. Utilizing some simplest wavelets, we demonstrate the competence of this framework in describing the baryon acoustic oscillation (BAO) data without any modifications to the agreement with cosmic microwave background measurements. This framework also provides a natural explanation for the bumps found in nonparametric observational reconstructions of the Hubble parameter and dark energy density as compensations of the dips suggested by some BAO data, and questions the physical reality of their existence. We note that utilizing this framework on top of the models that agree with both the cosmic microwave background and local $H_0$ measurements but are held back by BAO data, one may resurrect these models through the wiggly nature of wavelets that can naturally accommodate the BAO data. Finally, we also suggest narrowing the plausible set of admissible wavelets to further improve our framework by imposing conditions from expected kinematics of a viable cosmological model or first principle fundamental physics such as energy conditions.
\end{abstract}

\maketitle

\section{Introduction}
The base Lambda cold dark matter ($\Lambda$CDM) model is the simplest cosmological model that describes most of the data with remarkable accuracy~\cite{Riess:1998cb,SupernovaCosmologyProject:1998vns,Planck:2018vyg,Alam:2020sor,DES:2021wwk}. However, even if we set aside its persistent theoretical issues associated with the cosmological constant $\Lambda$~\cite{Weinberg:1988cp,Sahni:1999gb,Peebles:2002gy,Padmanabhan:2002ji}, with the increase in the diversity and precision of observational measurements and also with advances in data analysis and statistical methods, it has become increasingly plausible that a more realistic alternative model may be needed to replace the six-parameter base $\Lambda$CDM model as the new standard model of cosmology, and it seems that this new model should phenomenologically exhibit nontrivial/unexpected, if not significant, deviations from $\Lambda$CDM~\cite{DiValentino:2020vhf,DiValentino:2020zio,DiValentino:2020vvd,DiValentino:2020srs,DiValentino:2021izs,Perivolaropoulos:2021jda,Abdalla:2022yfr}. Some of these deviations, when associated with dark energy (DE) (as an effective or actual source), suggest phenomenological features that are difficult to obtain within the canonical/simple extensions of $\Lambda$CDM and are hard to deal with within the established fundamental theories of physics, for example, DE models that yield a density that attains negative values in the past which present at least one pole in their equation of state (EoS) parameters~\cite{Delubac:2014aqe,Aubourg:2014yra,Sahni:2014ooa,DiValentino:2017rcr,Mortsell:2018mfj,Poulin:2018zxs,Capozziello:2018jya,Wang:2018fng,Akarsu:2019ygx,Dutta:2018vmq,Banihashemi:2018oxo,Banihashemi:2018has,Farhang:2020sij,Banihashemi:2020wtb,Visinelli:2019qqu,Akarsu:2019hmw,Ye:2020btb,Ye:2020oix,Perez:2020cwa,Calderon:2020hoc,Paliathanasis:2020sfe,Bonilla:2020wbn,Vazquez:2012ag,Akarsu:2020yqa,LinaresCedeno:2021aqk,Zhou:2021xov,LinaresCedeno:2020uxx,DiValentino:2020naf,Akarsu:2021fol,Ozulker:2022slu,DiGennaro:2022ykp,Akarsu:2022typ,Malekjani:2023dky}, and/or present nontrivial characteristics such as an oscillatory EoS parameter that can even cross below the phantom divide line~\cite{Zhao:2017cud,Escamilla:2021uoj,Hojjati:2009ab,Xia:2006rr,Pace:2011kb,Cicoli:2018kdo,Ruchika:2020avj,Heisenberg:2022gqk,Tamayo:2019gqj,Lazkoz:2010gz,Pan:2017zoh}, and/or an oscillatory density~\cite{Colgain:2021pmf,Pogosian:2021mcs,Raveri:2021dbu,Wang:2018fng,Escamilla:2021uoj,Bernardo:2021cxi,Tamayo:2019gqj,Kazantzidis:2020xta}. This recent trend in cosmology is closely related to the fact that it is more challenging than originally thought to resolve the discordances (if not systematics) that emerge between different observations when assuming $\Lambda$CDM or its canonical/simple extensions. Although some of these discordances (e.g., the Ly-$\alpha$ anomaly) have decreased in significance with new probes, the fact that others have persisted (e.g., the $S_8$ tension), and some (e.g., the $H_0$ tension) have even increased in significance, lead an increasing number of researchers to think that these discordances cannot be attributed to unknown systematics. For a comprehensive reading on cosmological tensions and possible systematics in the data, we refer the reader to Refs.~\cite{Huterer:2017buf,Moresco:2022phi,DiValentino:2021izs,DiValentino:2022oon,Perivolaropoulos:2021jda,Abdalla:2022yfr} and references therein.

While such deviations are highly nontrivial, the reason they appear in some observational studies, as we will show, may be fairly simple. To begin with, let us describe the deviation of any alternative cosmological model from the $\Lambda$CDM model by ${\Delta H(z)\equiv H(z)-H_{\Lambda\rm CDM}(z)}$, where $H_{\Lambda\rm CDM}(z)$ is the Hubble function of the standard cosmological model, and $H(z)$ corresponds to the alternative model. Within the framework of the spatially flat Robertson-Walker (RW) metric, fixing the comoving angular diameter distance $D_M(z)=c\int_0^{z}\dd{z'}H^{-1}(z')$ of the alternative model to that of $\Lambda$CDM at any redshift $z=z_{\rm s}$, requires that $\int_0^{z_{\rm s}}\dd{z'}H^{-1}_{\Lambda \rm CDM}(z')= \int_0^{z_{\rm s}}\dd{z'}H^{-1}(z')$. This is satisfied only if there exist at least two redshifts $z_1$ and $z_2$ in the interval $(0,z_{\rm s})$ for which $\Delta H(z_1)/\Delta H(z_2)<0$ unless $\Delta H(z<z_{\rm s})$ vanishes everywhere. Thus, a negative $\Delta H(z)$ at any redshift (e.g., the apparent dip at $z\sim2.3$ if the Ly-$\alpha$ data is taken at face value~\cite{Sahni:2014ooa,Aubourg:2014yra,Planck:2018vyg}) should be compensated by at least one positive $\Delta H(z)$ somewhere else (e.g., the bump found in some DE density reconstructions at $1.5\lesssim z\lesssim2$~\cite{Wang:2018fng,Escamilla:2021uoj,Bernardo:2021cxi}). This compensatory behaviour implies an oscillation (not necessarily periodic) on top of $H_{\Lambda\rm CDM}(z)$ as suggested in the above mentioned observational analyses---which is in line with such behavior being favored by the baryon acoustic oscillations (BAO) data. An important consequence of this is that, due the same compensation, an observation with strict model independent constraints on $D_M(z_s)$ would render reconstructional approaches, or models with enough phenomenological flexibility, prone to finding artificial/fake bumps or dips due to overfitting, e.g., if the constraints on a model prefer a dip ($\Delta H<0$) to fit some data at a certain redshift better (for any measure of goodness of fit) than $\Lambda$CDM, its compensatory bump may arise as an artifact at redshifts where $H(z)$ is not directly constrained due to lack of data at those points; moreover, the preferred dip in our example may be due to overfitting, in which case both the dip and its compensatory bump would be fake. In particular, we can choose $z_{\rm s}$ to be the redshift of last scattering $z_*$ as $D_M(z_*)$ is strictly constrained by cosmic microwave background (CMB) observations almost model independently for a given prerecombination expansion history. This allows for oscillations up to $z_*$ with no constraints on their characteristics as long as they compensate each other so that the $D_M(z_*)$ integral is satisfied (they can be very frequent or just a single oscillation spread throughout the whole interval with an arbitrary shape etc.); however, it is conceivable that as the presently available cosmological observations other than CMB mostly probe redshifts $z\lesssim3$, the shape and the place of the oscillations will be constrained by these local data---nevertheless, these oscillations might have arisen as artifacts and/or been manipulated as noted above, in particular, due to overfitting.

In addition to the oscillations that arise from fixing the prerecombination Universe and hence $D_M(z_*)$ to that of $\Lambda$CDM, if one also respects the success of $\Lambda$CDM in the late Universe ($z\sim0$), we show that, the deviations from the Hubble radius, $H^{-1}(z)$, of $\Lambda$CDM are described by localized oscillatory functions, namely, \textit{wavelets}, and, these wavelets should satisfy the \textit{admissibility condition}. Admissible wavelets are oscillatory functions with a vanishing integral over their whole range, and either have compact support or vanish approximately outside of a compact set of their parameters~\cite{Chui:1992} (see the bottom panel of \cref{fig:hradius} for some wavelet examples). They can generically be obtained from derivatives of probability distributions but are by no means limited to this method. In cosmology, the wavelets have been used in various contexts. The wavelet transforms have been used in analyzing the CMB signals~\cite{Sanz:1999xu,Tenorio:1999mf,McEwen:2007ni}, and analyzing the large-scale structure of the Universe (to capture its non-Gaussian information content)~\cite{Einasto:2010bi,Allys:2020vld,Valogiannis:2021chp,Eickenberg:2022qvy,Wang:2021zse,Ajani:2020dvu,Ajani:2021pgp,Zurcher:2022clh,Ajani:2022ifk}; also, see Ref.~\cite{Valogiannis:2022xwu} and references therein for some applications of wavelets in cosmology and astrophysics. Wavelets have also been considered for investigating possible oscillatory deviations in the DE EoS parameter from minus unity describing the cosmological constant~\cite{Hojjati:2009ab}; however, note that, their approach of characterizing the oscillations of the EoS parameter with wavelets is fundamentally different from the central idea in this paper that deviations from $H^{-1}_{\Lambda \rm CDM}(z)$ must be described by admissible wavelets. Such deviations in the Hubble radius may also be described by wavelet oscillations in the EoS parameter, but, they may also correspond to much more violent behaviours with singularities, or even correspond to a cosmological constant if the wiggles in the Hubble radius are not attributed to the DE. Here, we show that models whose deviations from $\Lambda$CDM are described by admissible wavelets on top of $H^{-1}_{\Lambda \rm CDM}(z)$ constitute a family of cosmological models that are in excellent agreement with the CMB measurements; and, discuss how even the simplest wavelets can lead to nontrivial behaviours in the Hubble parameter that better describe the available BAO data without introducing an excessive number of free parameters. These deviations from $H^{-1}_{\Lambda\rm CDM}(z)$ can originate from different extensions in fundamental physics: modified theories of gravity, dynamical or nonminimally interacting DE etc. We discuss two such origins, i.e., dynamical DE and varying gravitational coupling strength; and, we expose through some simplest examples of wavelets, how the behaviour of some functions relevant to the source phenomena can be even more nontrivial. For example, if the deviations from $H^{-1}_{\Lambda\rm CDM}(z)$ are attributed to DE, the oscillations of the wavelet may cause the DE density to oscillate with a large enough amplitude so that the density attains negative values, resulting in divergences in its EoS parameter~\cite{Ozulker:2022slu}. Note that, while the same background dynamics may originate from different extensions, it may be possible to differentiate between these scenarios as we show by comparing the implications of attributing the deviations to the gravitational ``constant", with attributing them to the DE.\\

\vspace{-1cm}
This paper consists of three main parts. First, in \cref{sec:statement}, we put forward how admissible wavelets on top of $H^{-1}_{\Lambda{\rm CDM}}$ are mathematically implied under some observationally motivated conditions, and discuss their consequences on cosmological parameters, viz. bringing in wiggles on top of Hubble, deceleration and jerk parameters of the standard $\Lambda$CDM model. Then, in the second part (\cref{sec:de_descend,sec:g_descend}), we discuss the results of attributing these wavelets to different physical origins, such as the DE or gravitational coupling strength; and in the third part (\cref{sec:ex}), we demonstrate the potential implications that the wavelet modifications could have, by discussing the consequences of some simplest wavelet examples on various kinematical parameters and on the physical origin the wavelets are attributed to. Finally, we conclude in the last section.

\section{Wavelets on top of the standard cosmological model's Hubble radius}
\label{sec:statement}

We begin with the fact that the angular scale of the sound horizon at last scattering,
\be
\theta_*=\frac{r_*}{D_M(z_*)},\label{eq:theta}
\ee
is measured almost model independently, e.g., $100\theta_*=1.04110\pm 0.00031$ ($\Lambda$CDM Planck18~\cite{Planck:2018vyg}), with a precision of 0.03\%, where $r_*$ is the comoving sound horizon at last scattering, and $D_M(z_*)$ is the comoving angular diameter distance out to the last scattering surface. Then, fixing the prerecombination physics to that of the standard cosmological model, i.e., $\Lambda$CDM,
\be\label{eq:soundhorizon}
r_*=\int_{z_*}^\infty \frac{c_{\rm s}(z)}{H_{\Lambda \rm CDM}(z)}\dd{z}
\ee
is also determined, viz., $r_*=144.43\pm 0.26\,\rm Mpc$ ($\Lambda$CDM Planck18~\cite{Planck:2018vyg}). Here $c_{\rm s}(z)$ is the sound speed in the plasma and $z_{*}\approx 1090$ is the redshift of last scattering (redshift for which the optical depth to Thomson scattering reaches unity) and $H_{\Lambda \rm CDM}(z)$ is the Hubble parameter of the standard cosmological model: 
\be
3H^2_{\Lambda \rm CDM}(z)=\rho_{\rm m0}(1+z)^3+\rho_{\rm r0}(1+z)^4+\rho_{\Lambda},
\ee
where $\rho_{\rm m0}$, $\rho_{\rm r0}$, and $\rho_{\Lambda}$
are the present-day energy densities corresponding to those of the pressureless matter (m), the radiation (r), and the cosmological constant ($\Lambda$)---or, equivalently, the usual vacuum energy of the quantum field theory. We work, for convenience, in units for which the Newton's constant $G_{\rm N}=1/8\pi$ and the speed of light $c=1$ unless they are shown explicitly. Here, and in what follows, the subscript $0$ denotes the present-day ($z=0$) value of any quantity. While the values of these energy densities are subject to observational constraints, for the rest of this paper, we will assume them to be fixed (but unknown) values for which
\be
D_{M}(z_*)=c\int_0^{z_*}\frac{\dd{z}}{H_{\Lambda\rm CDM}(z)}
\ee
is consistent with~\cref{eq:theta,eq:soundhorizon}, and $\rho_{\rm m0}$ is compatible with the positions and relative heights of the peaks in the CMB power spectrum and $\rho_{\rm r0}$ is compatible with the observed CMB monopole temperature and standard model of particle physics. This ensures the basic consistency of $\Lambda$CDM with the CMB data at the background level.

Assuming $H_{\Lambda \rm CDM}(z)$ accurately describes the prerecombination universe (hence $r_*$ is known), for a universe described by the spatially flat RW metric, the comoving angular diameter distance to $z_*$,
\begin{equation}
      D_{M}(z_*)=c\int_0^{z_*}\frac{\dd{z}}{H(z)},\label{eq:DM}
\end{equation}
 of any model described by the Hubble parameter $H(z)$, is strictly constrained almost model independently through the measurement of $\theta_*$. Thus, for any $H(z)>0$ (expanding universe) deviating from $H_{\Lambda \rm CDM}(z)$, strict observational constraints from CMB still require
\be
\int_0^{z_*}\frac{\dd{z}}{H_{\Lambda \rm CDM}(z)}\approx \int_0^{z_*}\frac{\dd{z}}{H(z)}, \label{eq:approx}
\ee
cf., $D_M(z_*)=13872.83\pm 25.31\,\rm Mpc$ from $\Lambda$CDM Planck18~\cite{Planck:2018vyg}.
For simplicity, we will assume
the approximation in Eq.~\eqref{eq:approx} to be exact, and comment on the approximate case when necessary. Now, we define the deviation of a cosmological model from $\Lambda$CDM in terms of its Hubble radius, $H(z)^{-1}$, as follows:
\be
\psi(z)\equiv\frac{1}{H(z)}-\frac{1}{H_{\Lambda \rm CDM}(z)}.\label{eqn:devdef}
\ee
Then, we have
\be
 D_{M}(z_*)=c\int_0^{z_*}\dd{z}\qty[\frac{1}{H_{\Lambda \rm CDM}(z)}+\psi(z)],\label{eq:exact}
\ee
and consequently, the exact version of Eq.~\eqref{eq:approx} implies
\be
\Psi(z_*)\equiv\int_0^{z_*}\psi(z) \dd{z}=0.\label{eq:int}
\ee
Our assumption that the prerecombination universe is accurately described by $H_{\Lambda \rm CDM}(z)$, viz., $H(z\geq z_*)=H_{\Lambda \rm CDM}(z\geq z_*)$, implies another condition on $\psi(z)$, that is, 
\be
\psi(z\geq z_*)=0. \label{eq:prereccond}
\ee

This mathematical framework allows one to naturally classify a family of $H(z)$ functions which can deviate, even significantly, from $H_{\Lambda \rm CDM}(z)$, but still have the same $D_M(z_*)$ the $\Lambda$CDM model has, ensuring basic consistency with the CMB measurements at the background level (one might want to also consider the constraints on $\rho_{\rm m0}$ and $\rho_{\rm r0}$ from CMB). This family is described by 
\be
H(z)=\frac{H_{\Lambda \rm CDM}(z)}{1+\psi(z)H_{\Lambda \rm CDM}(z)},\label{eq:hz}
\ee
where $\psi(z)$ satisfies the conditions introduced in~\cref{eq:int,eq:prereccond}. We notice from this equation that introduction of the condition $-H^{-1}_{\Lambda \rm CDM}(z)<\psi(z)<\infty$ ensures that, in the past ($z>0$), $H(z)$ never diverges (except at the big bang) and the Universe has always been expanding. Also, on top of all these conditions, let us demand 
\be
\psi(z=0)=0\label{eq:present}
\ee
since we know the Universe at $z\sim0$ is well described by the standard $\Lambda$CDM model~\cite{Planck:2018vyg,Alam:2020sor,DES:2021wwk,Colgain:2021pmf}.

We notice that \cref{eq:int,eq:prereccond,eq:present} describe characteristic properties of functions that are known as \textit{wavelets} where \cref{eq:int} is true for wavelets that satisfy the \textit{admissibility condition}~\cite{Chui:1992}. Wavelets are oscillatory (not necessarily periodic) functions that are well localized, i.e., they have compact support or they vanish approximately outside of a compact set of its parameters (see the bottom panel of \cref{fig:hradius} for some wavelet examples). With such boundary conditions that the function should absolutely or approximately vanish outside of certain bounds, \cref{eq:int} requires that the function oscillates at least once if it does not vanish everywhere; because, say $\psi(z)<0$ for a certain value of $z$, this integral can vanish only if $\psi(z)>0$ at another value of $z$, hence the oscillation. Note that, for a continuous $\psi(z)$, this argument also implies that there exists at least one value of $z$ in the interval $(0,z_*)$ for which $\psi=0$; this corresponds to the Rolle's theorem, which, in our particular case, states that the conditions $\Psi(0)=0$ and $\Psi(z_*)=0$ imply the existence of a $z_p\in(0,z_*)$ for which $\psi(z_p)=0$. Thus, the deviations from the standard $\Lambda$CDM model's Hubble radius, $\psi(z)$, must be described by admissible wavelets, i.e., must have a wiggly (wavelike) behaviour characterized by the conditions given in \cref{eq:int,eq:prereccond,eq:present}.

We proceed with showing explicitly that the characteristics of $\psi(z)$ described above corresponds to a wiggly behaviour for $H(z)$ with respect to $H_{\Lambda \rm CDM}(z)$ \textit{in a particular way}; to see this, we define a unitless parameter $\delta(z)$, namely, the fractional deviation from $H_{\Lambda\rm CDM}(z)$, as follows:
\be
\label{eqn:deltaH}
\delta(z)\equiv \frac{H(z)-H_{\Lambda \rm CDM}(z)}{H_{\Lambda \rm CDM}(z)}=-\frac{\psi(z)H_{\Lambda \rm CDM}(z)}{1+\psi(z)H_{\Lambda \rm CDM}(z)}.
\ee
We see that if we demand an ever-expanding universe $H(z)>0$, we should set $\delta(z)>-1$. And, in what follows, unless otherwise is stated, we continue our discussions with the assumption that $\delta(z)>-1$. For small deviations from $\Lambda$CDM, i.e., $|\delta(z)|\ll1$, we can also write
\be
\label{eqn:sdevH}
\delta(z)\approx-\psi(z)H_{\Lambda \rm CDM}(z).
\ee
The small deviation region is quite important to study, because, despite its shortcomings, $\Lambda$CDM is still the simplest model to explain the cosmological observations with remarkable accuracy. Particularly, in the late universe, the small deviation approximation is robustly imposed by many cosmological probes that require $|\delta(z)|\ll1$ for ${z\lesssim2.5}$; even the largest discrepancies between the $H_{\Lambda \rm CDM}(z)$ of the Planck 2018 $\Lambda$CDM~\cite{Planck:2018vyg} and observed $H(z)$ values, viz., $H_0=73.04 \pm 1.04$ km s${}^{-1}$ Mpc${}^{-1}$ (the SH0ES $H_0$ measurement~\cite{Riess:2021jrx}) and $H(2.33)=224\pm8{\rm \,km\, s^{-1}\, Mpc^{-1}}$ (the Ly-$\alpha$-quasar data~\cite{eBOSS:2020yzd}) correspond to $\delta_0\sim0.08$ and $\delta(z=2.33)\sim-0.05$, respectively. The form of~\cref{eqn:sdevH} makes it even easier to see that $H(z)$ will have wiggles; since $H_{\Lambda \rm CDM}(z)$ is a monotonically varying function of $z$ and strictly positive, when $\psi(z)$ changes sign (as it must at least once), this sign change [around which the small deviation condition is clearly satisfied for continuous $\psi(z)$] is directly reflected on $\delta(z)$, producing a wiggle. Furthermore, respecting the successes of the $\Lambda$CDM model, one may even wish to impose $|\delta(z)|\ll1$ at all times. In this case, since $H_{\Lambda \rm CDM}(z)$ monotonically grows with increasing redshift, one should have $\psi(z)\to0$ fast enough with $z\to z_*$, such that the small deviation condition $\abs{\psi(z)H_{\Lambda \rm CDM}(z)}\ll1$ is not broken.

Having said that, note the interesting extra behaviours apparent from the full form of $\delta(z)$ in~\cref{eqn:deltaH}: first, as mentioned before, ${\psi(z)=-H^{-1}_{\Lambda \rm CDM}(z)}$ results in a singular $H(z)$ function and is not allowed for finite $z$ values; second, while the previous condition might seem to require either one of the confinements ${\psi(z)>-H^{-1}_{\Lambda \rm CDM}(z)}$ or ${\psi(z)<-H^{-1}_{\Lambda \rm CDM}(z)}$ at all times, in principle, $\psi(z)$ can be discontinuous and is not necessarily confined to one of these regions; third, \cref{eqn:sdevH} indicates that $\psi(z)<0$ corresponds to $\delta(z)>0$, yet, for a region in which ${\psi(z)<-H^{-1}_{\Lambda \rm CDM}(z)}$, we have $\delta(z)<0$ despite having $\psi(z)<0$, but, looking at~\cref{eqn:devdef}, such a region also corresponds to an extreme case with $H(z)<0$ and the Universe would have gone through a contracting phase.

Finally, it is worth noting that due to the wiggly behaviour of the wavelets, similar to $H(z)$, the other important kinematical parameters in cosmology, the deceleration parameter $q=-1+\frac{{\rm d}}{{\rm d}t}\left[H^{-1}(z)\right]$ (where $t$ is the cosmic time) and the jerk parameter $j=\frac{{\rm d}^3a/{\rm d}t^3}{aH^3(z)}$ (which is simply $j_{\Lambda\rm CDM}=1$ for $\Lambda$CDM) will also exhibit wiggly behaviors; the deceleration parameter will oscillate around its usual evolution in $\Lambda$CDM, $q_{\Lambda\rm CDM}(z)$, as can be immediately seen from
\begin{equation}
    q(z)=q_{\Lambda\rm CDM}(z)+\frac{{\rm d}\psi(z)}{{\rm d}t},
\end{equation}
obtained by using \cref{eqn:devdef} in the definition of $q(z)$. And, the jerk parameter will oscillate around its constant $\Lambda$CDM value of unity. These behaviours are reminiscent of the nonparametric reconstructions in Refs.~\cite{Mukherjee:2020vkx,Mukherjee:2020ytg}.

\section{Wiggles in dark energy density descended from the wavelets}
\label{sec:de_descend}

In the late universe where dust and DE are the only relevant components, we can treat $H(z)$ as an extension of ${H_{\Lambda \rm CDM}}(z)$ with the same matter density parameter $\rho_{\rm m}(z)$ but with a minimally interacting dynamical DE that explains the deviation of $\delta(z)$ from zero; hereby, we can write the DE density as $\rho_{\rm DE}(z)\equiv3H^2(z)-\rho_{\rm m}(z)$, viz.,
\be
\begin{aligned}
\label{eqn:drho}
\rho_{\rm DE}(z)&=3H^2_{\Lambda \rm CDM}(z)[1+\delta(z)]^2-\rho_{\rm m0}\qty(1+z)^3\\
&=\rho_{\rm DE0}+3H^2_{\Lambda \rm CDM}(z)\delta(z)[2+\delta(z)],
\end{aligned}
\ee
from which we can write the deviation of the DE density from $\Lambda$, i.e., ${\Delta\rho_{\rm DE}(z)\equiv\rho_{\rm DE}(z)-\rho_{\Lambda}}$ (where we have ${\rho_\Lambda=\rho_{\rm DE0}}$), as follows:
\be
\Delta\rho_{\rm DE}(z)=3H^2_{\Lambda \rm CDM}(z)\delta(z)[2+\delta(z)].\label{eq:deltarho}
\ee
For small deviations from $\Lambda$CDM, these read
\begin{align}
 \rho_{\rm DE}(z)&\approx \rho_{\rm DE0}+6\delta(z)H^2_{\Lambda \rm CDM}(z),\\
 \Delta\rho_{\rm DE}(z)&\approx6\delta(z)H^2_{\Lambda \rm CDM}(z),
\end{align}
correspondingly. Thus, because $\delta(z)$ is oscillatory around zero, $\Delta\rho_{\rm DE}(z)$ will also be oscillatory around zero and this oscillatory $\Delta\rho_{\rm DE}(z)$ corresponds to the oscillatory $\delta(z)$ scaled by $6H^2_{\Lambda \rm CDM}(z)$. That is, observational fitting/nonparametric reconstruction procedures predicting wiggles in $H(z)$ will predict corresponding wiggles in $\rho_{\rm DE}(z)$ reconstructions.

Even if our assumption that the prerecombination universe is not modified with respect to the standard cosmology [implying~\cref{eq:prereccond}], is taken to be approximate, for $z>z_*$, the fluctuations in the DE density should be much smaller than the matter energy density, ${\abs{\Delta\rho_{\rm DE}(z)/\rho_{\rm m}(z)}\ll1}$, in the matter dominated epoch, and much smaller than the radiation energy density, ${\abs{\Delta\rho_{\rm DE}(z)/\rho_{\rm r}(z)}\ll1}$, in the radiation dominated epoch. Since for both of these epochs the relevant energy densities can be well approximated by the critical energy density of $\Lambda$CDM, $\rho_{\rm c}(z)\equiv3H^2_{\Lambda\rm CDM}(z)$, in this approximate case for $z>z_*$, instead of $\Delta \rho_{\rm DE}(z)=0$, we can write the more relaxed condition
\be
 \abs{\frac{\Delta\rho_{\rm DE}(z)}{\rho_{\rm c}(z)}}=\abs{\delta(z)[2+\delta(z)]}\ll1. \label{eq:pert}
\ee
This is satisfied for both $\delta(z)\sim 0$ (small deviation from $\Lambda$CDM), and $\delta(z)\sim-2$ (corresponds to a contracting universe), but only the former is of interest to us. Since~\cref{eq:pert} requires small $|\delta(z)|$ to be satisfied, it can be rewritten as
\be
\abs{\frac{\Delta\rho_{\rm DE}(z)}{\rho_{\rm c}(z)}}\approx2\abs{\delta(z)}\approx2\abs{-\psi(z)H_{\Lambda \rm CDM}(z)}\ll1, \label{eq:rapidity}
\ee
from which we immediately see that $\psi(z)$ should vanish rapidly enough with increasing $z$ at large redshifts so that our assumption of almost unmodified prerecombination physics holds.

We calculate from Eq.~\eqref{eqn:drho} that the DE density passes below zero, $\rho_{\rm DE}(z)<0$, for
\be
\delta(z)<-1+\sqrt{1-\frac{\rho_{\rm DE0}}{3H^2_{\Lambda \rm CDM}(z)}},
\ee
which can also be written as follows:
\be
\delta(z)<-1+\sqrt{1-\frac{\Omega_{\rm DE0}}{\Omega_{\rm DE0}+(1-\Omega_{\rm DE0})(1+z)^3}}.
\ee
Accordingly, using Planck 2018 best fit $\Lambda$CDM values $\Omega_{\rm m0}=0.3158$ and $H_0=67.32{\rm \,km\, s^{-1}\, Mpc^{-1}}$~\cite{Planck:2018vyg}, it turns out that $\delta (2.33)<-0.028$, i.e., 
$\Delta H(2.33)\equiv{H(2.33)-H_{\Lambda\rm CDM}(2.33)<-6.65{\rm \,km\, s^{-1}\, Mpc^{-1}}}$ (corresponding to $H(2.33)\lesssim230.536{\rm \,km\, s^{-1}\, Mpc^{-1}}$), requires the DE density to yield negative values. Note that $H(2.33)=228\pm7{\rm \,km\, s^{-1}\, Mpc^{-1}}$ from the Ly-$\alpha$-Ly-$\alpha$ and $H(2.33)=224\pm8{\rm \,km\, s^{-1}\, Mpc^{-1}}$ from the Ly-$\alpha$-quasar data~\cite{eBOSS:2020yzd}. Thus, if all deviations from Planck 2018 best fit $\Lambda$CDM are attributed to the DE, these data are consistent with vanishing/negative DE density at $z\sim2.3$ when their $1\sigma$ error bars are considered and prefer a negative DE density at $z\sim2.3$ when their mean values are considered. In this sense, the function $\delta(z)$ can also be used as a diagnostic to test for a negative DE density if all modifications to Planck $\Lambda$CDM are attributed to the DE.

Lastly, the continuity equation for the DE, viz., $\dot\rho_{\rm DE}(z)+3H(z)[\rho_{\rm DE}(z)+p_{\rm DE}(z)]=0$, implies $\varrho_{\rm DE}(z)=\frac{1+z}{3} \rho_{\rm DE}'(z)$ for the inertial mass density, $\varrho_{\rm DE}(z)\equiv\rho_{\rm DE}(z)+p_{\rm DE}(z)$, and $w_{\rm DE}(z)=-1+\frac{1+z}{3} \rho_{\rm DE}'(z)/\rho_{\rm DE}(z)$ for the corresponding EoS parameter $w_{\rm DE}(z)\equiv p_{\rm DE}(z)/\rho_{\rm DE}(z)$, where $\rho_{\rm DE}(z)$ is the DE density as defined in~\eqref{eqn:drho}, $p_{\rm DE}(z)$ is its pressure, and $'\equiv\dv{z}$. Accordingly, we have
\be
\begin{aligned}
    \varrho_{\rm DE}(z)&=2(1+z)H_{\Lambda\rm CDM}^2\qty[\frac{H'_{\Lambda\rm CDM}}{H_{\Lambda\rm CDM}}\delta(\delta+2)+\delta'(\delta+1)]\\
    &\approx2(1+z)H^2_{\Lambda\rm CDM}\qty[2\frac{H'_{\Lambda\rm CDM}}{H_{\Lambda\rm CDM}}\delta+\delta'],
\end{aligned}
\ee
for the DE inertial mass density, and
\be
\begin{aligned}\label{eq:eos}
    w_{\rm DE}(z)&=-1+\frac{2(1+z)\qty[\frac{H'_{\Lambda\rm CDM}}{H_{\Lambda\rm CDM}}\delta(\delta+2)+\delta'(\delta+1)]}{3\qty[\frac{\rho_{\rm DE0}}{\rho_{\rm c}}+\delta(2+\delta)]}\\
    &\approx-1+\frac{2(1+z)\qty[2\frac{H'_{\Lambda\rm CDM}}{H_{\Lambda\rm CDM}}\delta+\delta']}{3\qty[\frac{\rho_{\rm DE0}}{\rho_{\rm c}}+2\delta]},
\end{aligned}
\ee
for the corresponding DE EoS parameter; in these two equations, the second lines are for small deviations from $\Lambda$CDM. Notice that, in the exact form of~\cref{eq:eos}, $w_{\rm DE}(z)$ blows up if $\rho_{\rm DE0}/\rho_{\rm c}(z)=-\delta(z)[2+\delta(z)]$ is satisfied for a redshift, say, at $z=z_{\rm v}$. Comparing with \cref{eqn:drho}, we see that this condition is equivalent to $\rho_{\rm DE}(z_{\rm v})=0$; indeed, if the DE submits to the continuity equation as it does in this case, a vanishing energy density necessitates such a singularity~\cite{Ozulker:2022slu}. Such infinities in the EoS parameter are not problematic from the fundamental physics point of view, instead, hints that the DE density is perhaps an effective one originating from a modified gravity model.

\section{Wiggles in Newton's ``constant" descended from the wavelets}
\label{sec:g_descend}

Alternatively, we can attribute the deviation of $H(z)$ from $H_{\Lambda \rm CDM}(z)$ to the deviations in the gravitational coupling strength, $G_{\rm eff}(z)$, from the Newton's gravitational constant $G_{\rm N}$ measured locally. We have, as usual,
\be
3H^2_{\Lambda \rm CDM}(z)=8\pi G_{\rm N}\qty[\rho_{\rm m0}(1+z)^3+\rho_{\rm r0}(1+z)^4+\rho_\Lambda],
\ee
where the constant value $\rho_\Lambda$ is either the usual vacuum energy density or $\rho_\Lambda=\frac{\Lambda}{8\pi G_{\rm N}}$. We can write the Hubble parameter of the new model as
\be
3H^2(z)=8\pi G_{\rm eff}(z)\qty[\rho_{\rm m0}(1+z)^3+\rho_{\rm r0}(1+z)^4+\rho_\Lambda],
\ee
from which, using the definition in~\cref{eqn:deltaH},
\be
\begin{aligned}
\label{eqn:Geff}
G_{\rm eff}(z)=\qty[1+\delta(z)]^{2}G_{\rm N}
\end{aligned}
\ee
directly follows. Note that $G_{\rm eff}(z)$ is also a wiggly function led by the wiggles of $\delta(z)$, but $G_{\rm eff}(z)$ equals $G_{\rm N}$ when $\psi(z)=0$, and thereby, $G_{\rm eff}(z=0)=G_{\rm eff}(z>z_*)=G_{\rm N}$ from~\cref{eq:prereccond,eq:present}. And,
for small deviations from $\Lambda$CDM, \cref{eqn:Geff} reads
\be
\begin{aligned}
G_{\rm eff}(z)\approx [1+2\delta(z)]G_{\rm N}.
\end{aligned}
\ee
Note that, if we are to treat $\rho_\Lambda$ as the effective energy density of the cosmological ``constant", i.e., ${\Tilde{\Lambda}(z)=8\pi G_{\rm eff}(z) \rho_{\Lambda}}$, this new cosmological term $\Tilde{\Lambda}(z)$ is not a constant anymore.

It is crucial to note that, while attributing the wiggles to the DE density or $G_{\rm eff}(z)$ is indistinguishable in their background dynamics, this is not so for all physical observables. Particularly, a direct effect of the dynamical gravitational coupling strength would be observable, for instance, as this would promote the absolute magnitude $M_B=\rm const$ of type Ia supernovae (SNIa) to a quantity varying with the redshift $M_B=M_B(z)$. Such an effect in the very late universe ($z\lesssim0.1$) was recently suggested and investigated in a series of papers to address the so-called $M_B$ (and $H_0$) tension~\cite{Alestas:2020zol,Marra:2021fvf,Perivolaropoulos:2021bds,Alestas:2022xxm,Perivolaropoulos:2022vql,Perivolaropoulos:2022txg}. Also, the idea that the supernovae absolute magnitudes are constant with redshift, has been questioned by observations and the question of whether or not this idea is valid has recently gained interest~\cite{Benisty:2022psx,DiValentino:2020evt,Rose:2020shp,Kang:2019azh,Kim:2019npy,Tutusaus:2017ibk,Linden:2009vh,Ferramacho:2008ap}. A possible variation of the $M_B(z)$ and equivalently of the SNIa luminosity $L(z)\propto 10^{-\frac{2}{5}M_B(z)}$ could be due to a variation of the Newton's ``constant". Since the SNIa luminosity is proportional to the Chandrasekhar mass, which, in this case, is no longer a constant equal to $1.4\,M_{\odot}$, but a quantity that varies with $G_{\rm eff}(z)$, we have $L(z)\propto M_{\rm Chandra}(z)$, so that
$L(z)\propto G_{\rm eff}^{-3/2}(z)$, which in turn leads, in this approach, to
\be
\begin{aligned}
M_{B}(z)-M_{B,G_{\rm N}}=\frac{15}{4}\log \frac{G_{\rm eff}(z)}{G_{\rm N}}=\frac{15}{2}\log[1+\delta(z)],\label{eq:mbwig}
\end{aligned}
\ee
where $M_{B,G_{\rm N}}$ denotes the SNIa absolute magnitude when $G_{\rm eff}(z)=G_{\rm N}$, which satisfies ${M_{B,G_{\rm N}}=M_{B,0}}$ due to~\cref{eq:present}.
Thus, attributing wiggles to $G_{\rm eff}(z)$ will have consequences not only on the expansion of the Universe, but also on the absolute magnitudes of SNIa at different redshifts; and, as \cref{eq:mbwig} shows, the wiggles of $G_{\rm eff}(z)$ are directly manifested in the SNIa absolute magnitudes as a wiggly $M_B(z)$ reminiscent of the findings of Ref.~\cite{Benisty:2022psx}. Investigating how this dual modification to the standard cosmology affects the cosmological parameter estimates from SNIa data and furthermore the so-called $M_B$ tension~\cite{Efstathiou:2021ocp,Camarena:2021jlr}, is beyond the scope of this paper, and deserves a separate study.

\begin{figure}[t!]
    \begin{flushright}
    \includegraphics[width=0.447\textwidth]{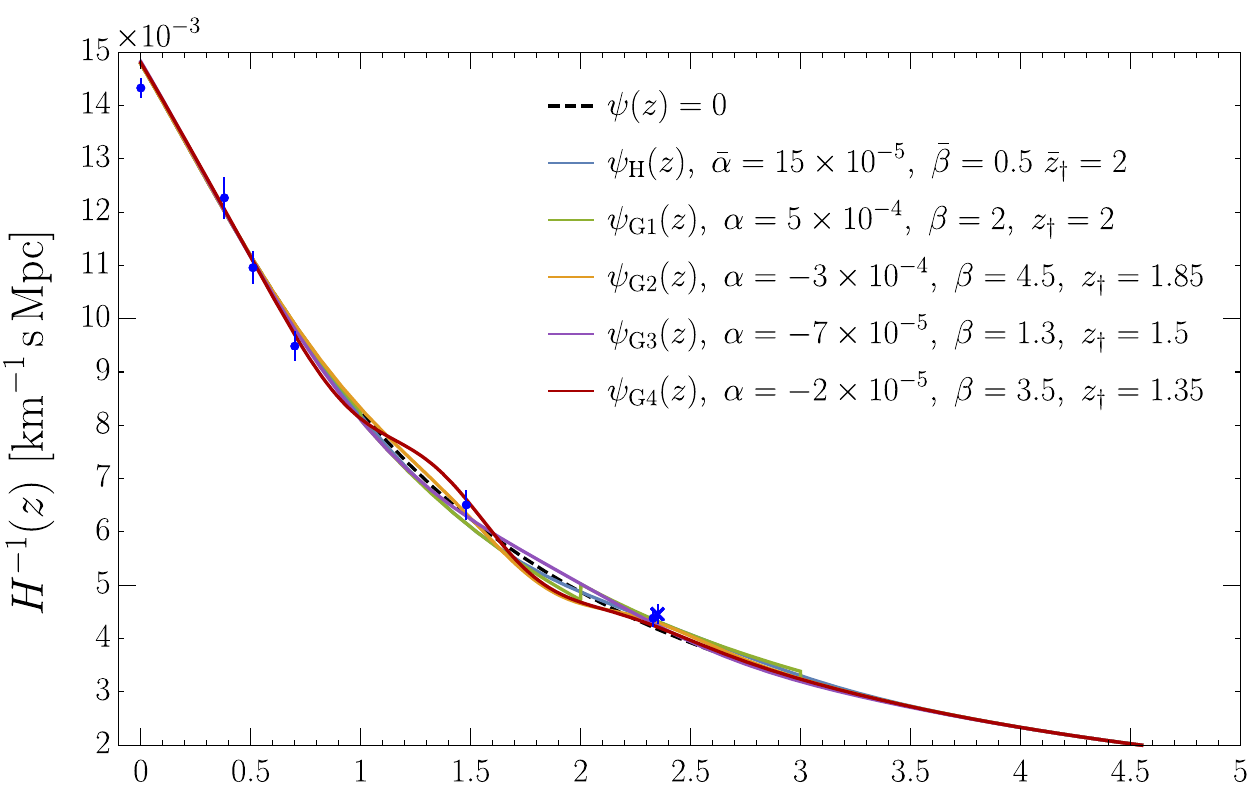}
    \includegraphics[width=0.45\textwidth]{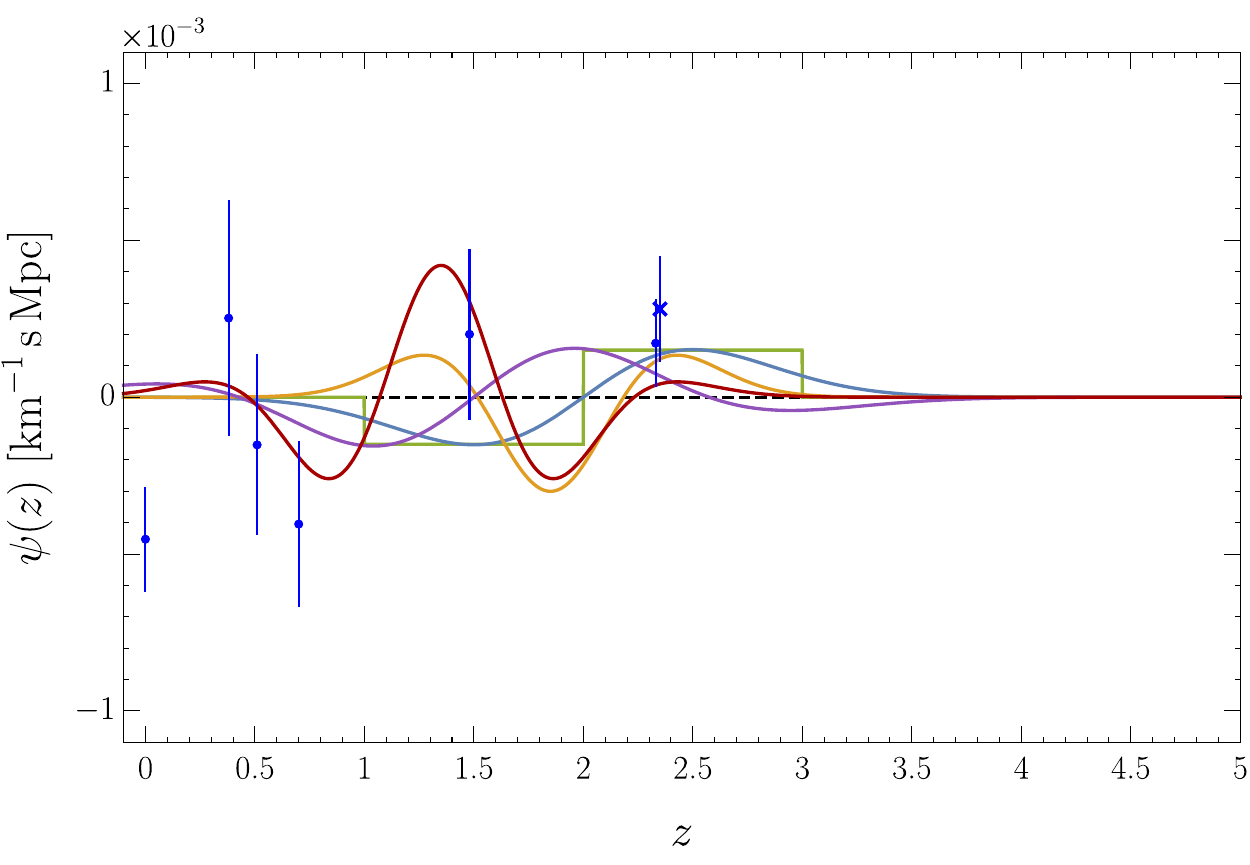}
    \end{flushright}
    \caption{The top panel shows the Hubble radii for some wavelet examples of $\psi(z)$ given in the bottom panel where $\Bar{\alpha}$ and $\alpha$ are in units of ${\rm \,km\, s^{-1}\, Mpc^{-1}}$, $\Bar{\beta}$ and $\beta$ are unitless, and $\Bar{z}_\dagger$ and $z_\dagger$ are redshifts anchoring the wavelets. The dashed line, $\psi(z)=0$, corresponds to no deviation, i.e., $\Lambda$CDM itself. The blue bars correspond to the TRGB $H_0$ measurement and various BAO measurements. See \cref{sec:ex} for details.
    }
    \label{fig:hradius}
\end{figure}

\section{Employing some simplest wavelets}
\label{sec:ex}

\begin{figure*}[ht!]
    \centering
    \makebox[0.49\textwidth][r]{\includegraphics[width=0.48\textwidth]{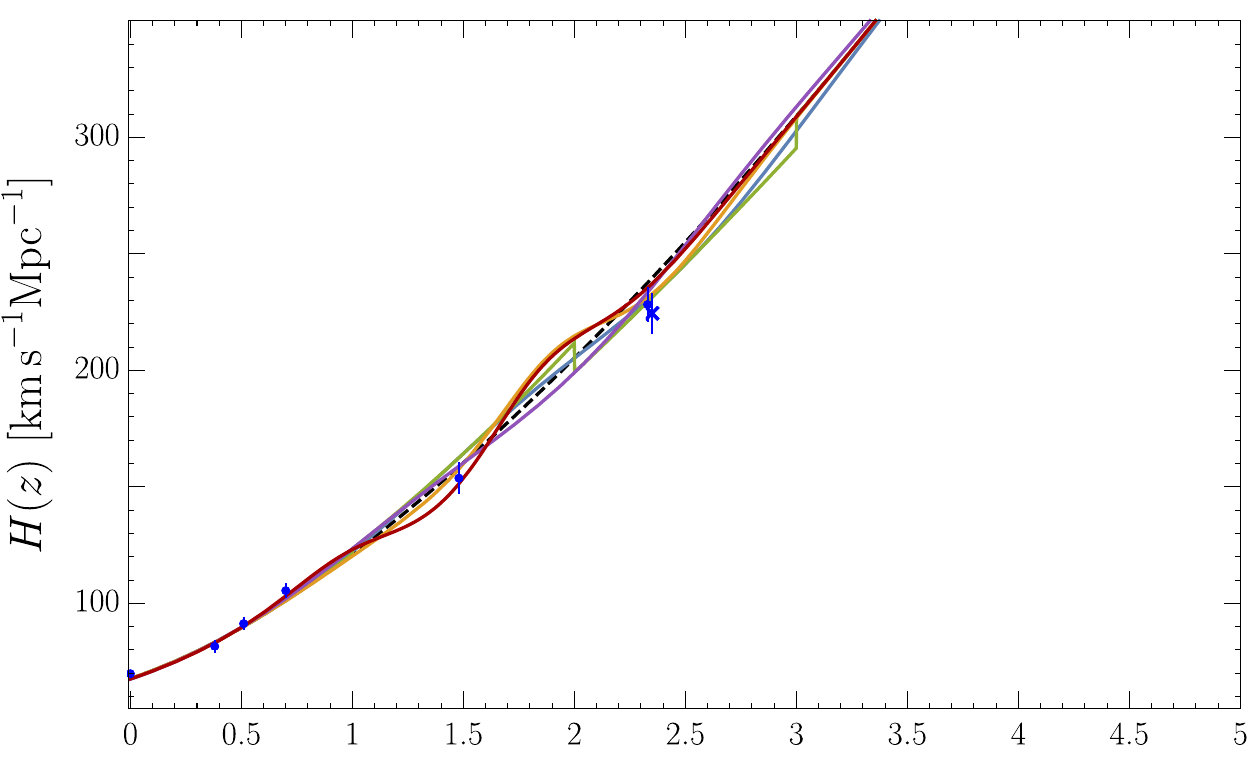}}
    \makebox[0.49\textwidth][r]{\includegraphics[width=0.48\textwidth]{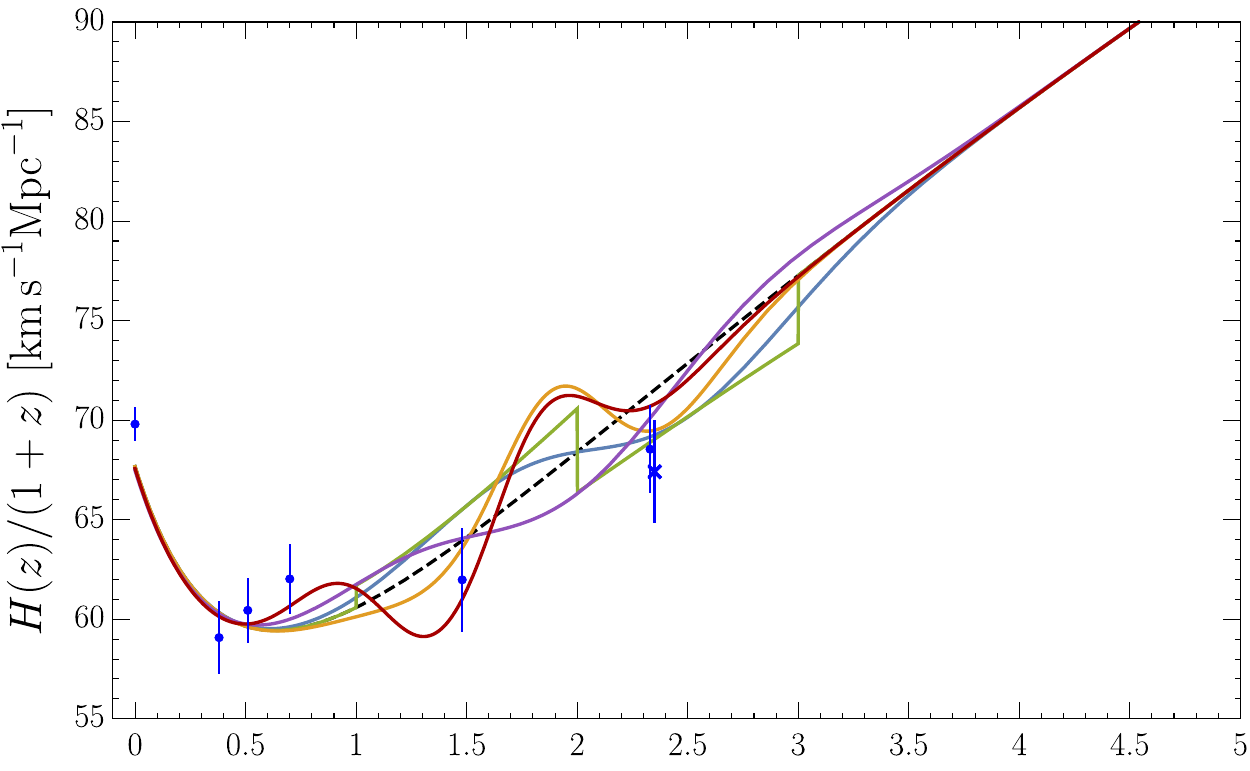}}
    
      \makebox[0.49\textwidth][r]{\includegraphics[width=0.48\textwidth]{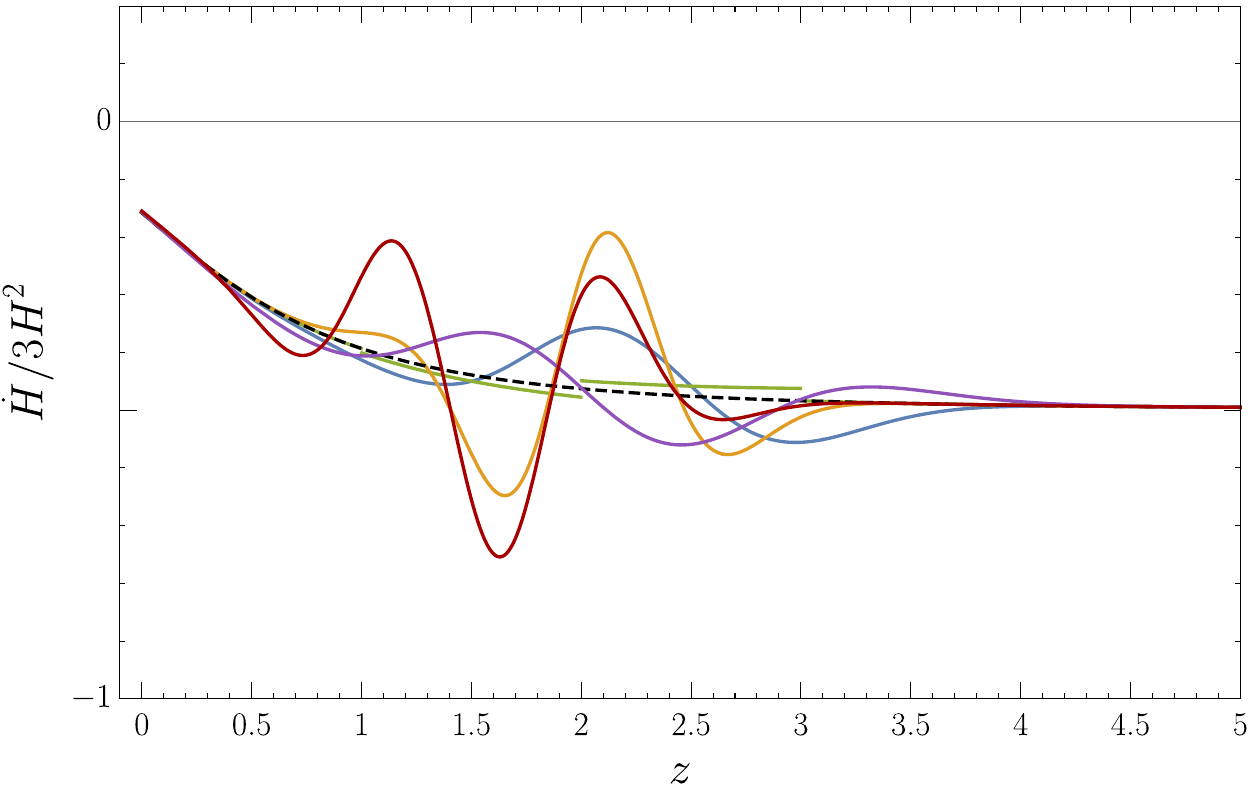}}
      \makebox[0.49\textwidth][r]{\includegraphics[width=0.48\textwidth]{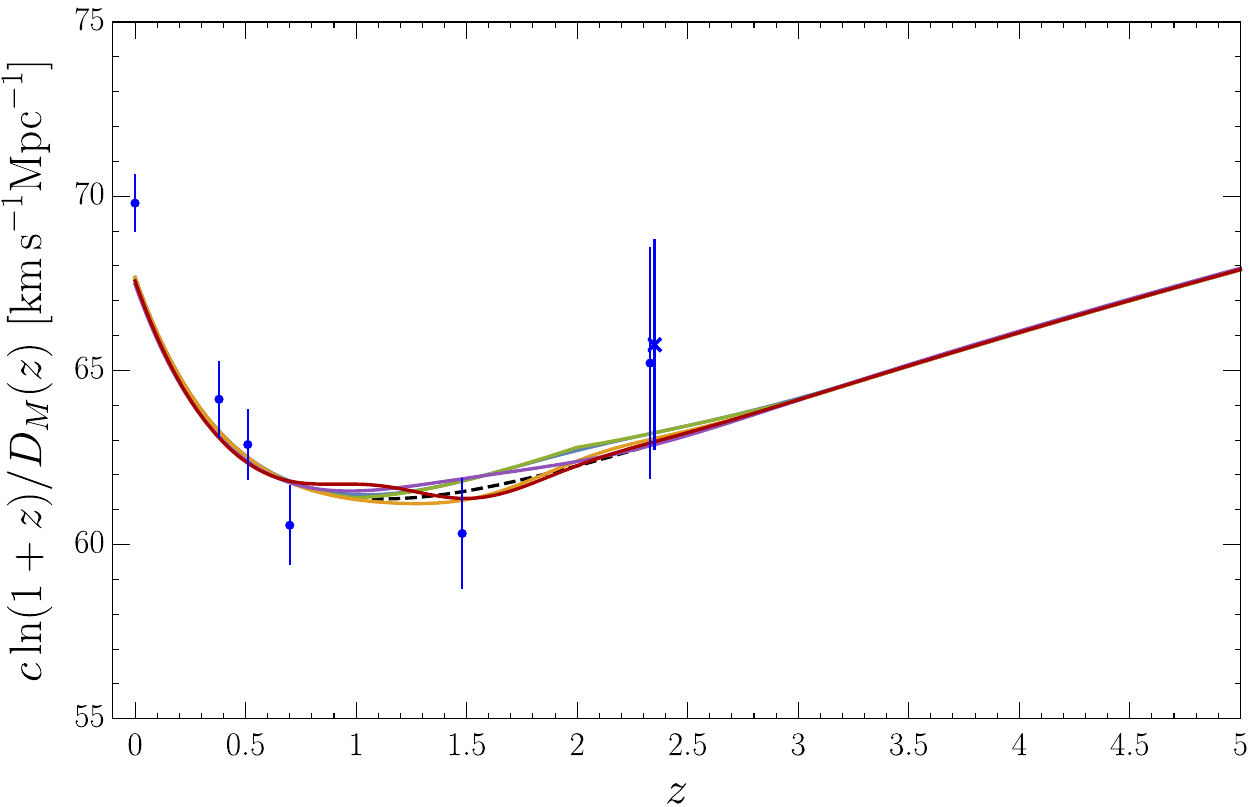}}
      
    \caption{The deviations from the $\Lambda$CDM model in terms of some kinematical parameters for the wavelet examples in \cref{fig:hradius}; the plots are matched by color to those in \cref{fig:hradius}}
    \label{fig:kinematics}
\end{figure*}

Wavelets constitute a wide family of functions that may or may not be smooth. They exhibit an oscillatory (not necessarily periodic) behaviour over a compact set of their parameters, and either vanish or quickly decay outside of this set. Even the superposition of arbitrarily many wavelets would describe another one. Here, we will consider some of the simplest examples: one discontinuous, namely, the Haar mother wavelet (\cref{sec:haar}); and other smooth wavelets, namely, the Hermitian wavelets (\cref{sec:hermitian}) that are acquired from the derivative/s of a Gaussian distribution function. These examples have no inherent superiority to other possible wavelets; we provide them only because of their simplicity and to give a taste of how wavelets behave and their cosmological consequences.

\begin{figure*}[ht!]
    \centering
    \includegraphics[width=0.45\textwidth]{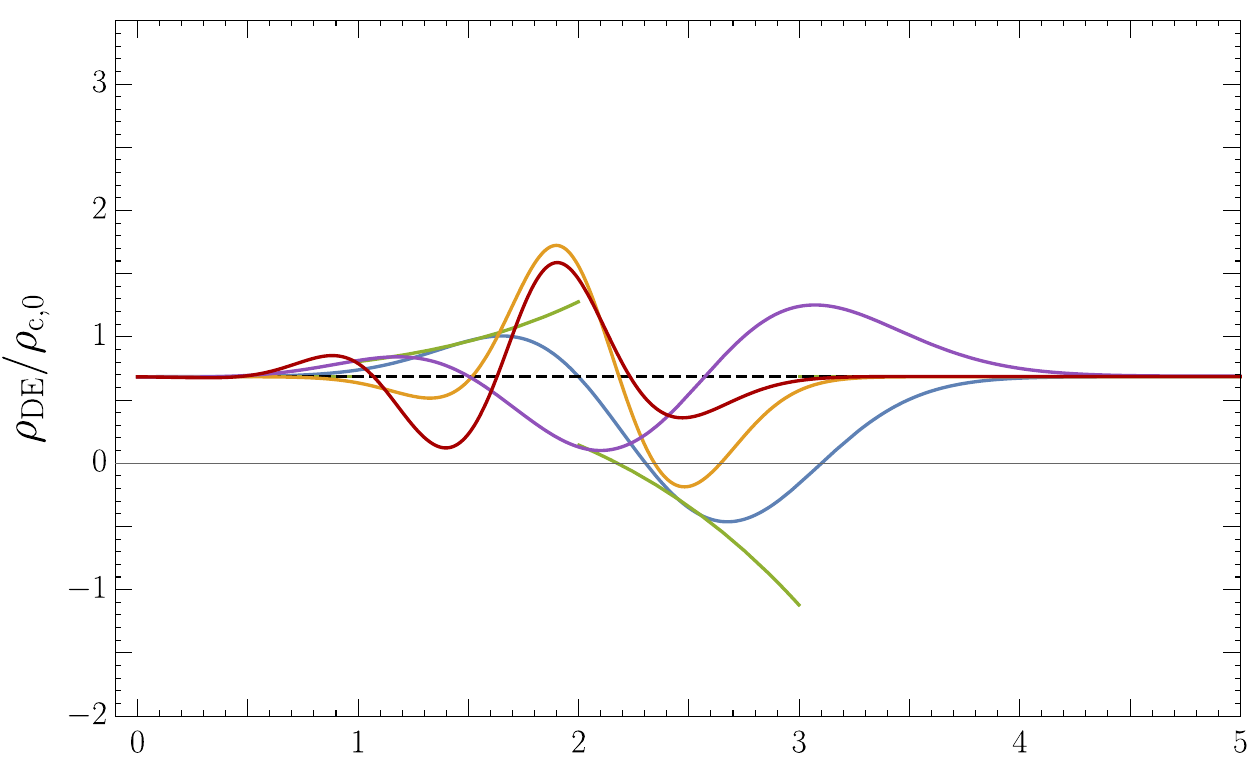}
    \includegraphics[width=0.45\textwidth]{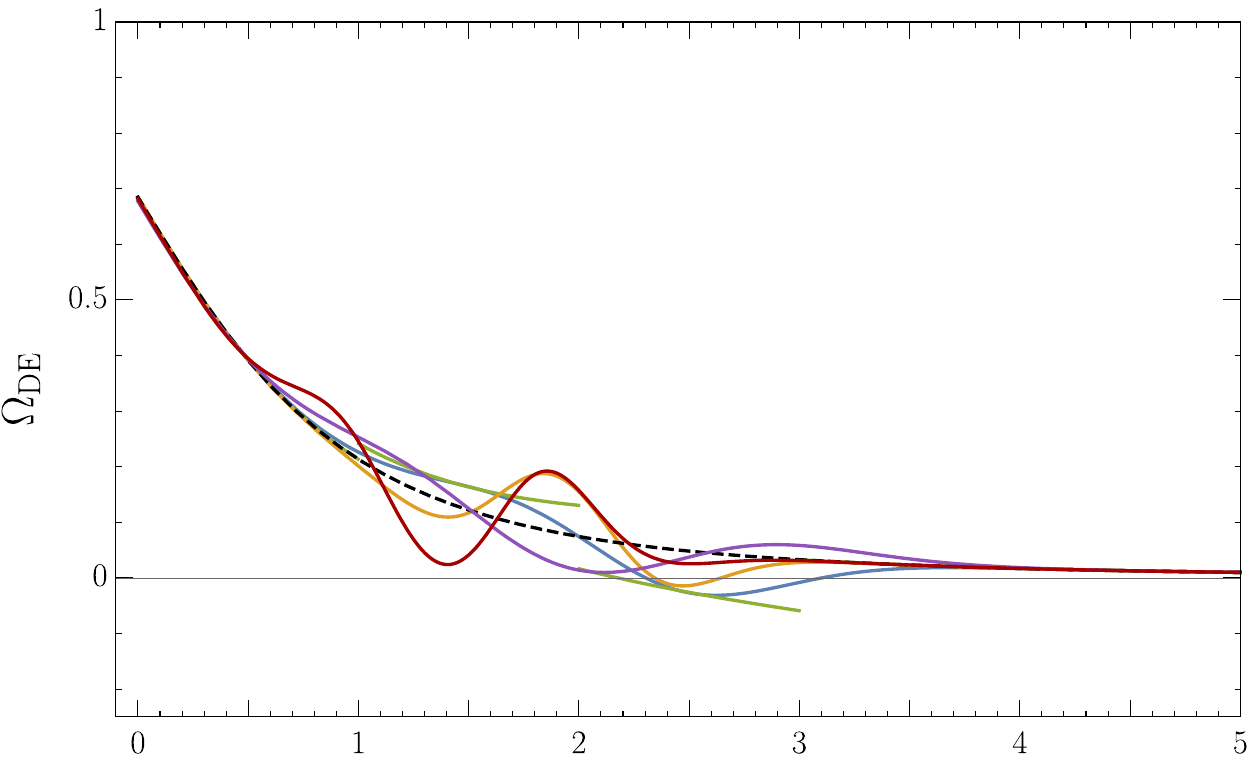}
    \includegraphics[width=0.45\textwidth]{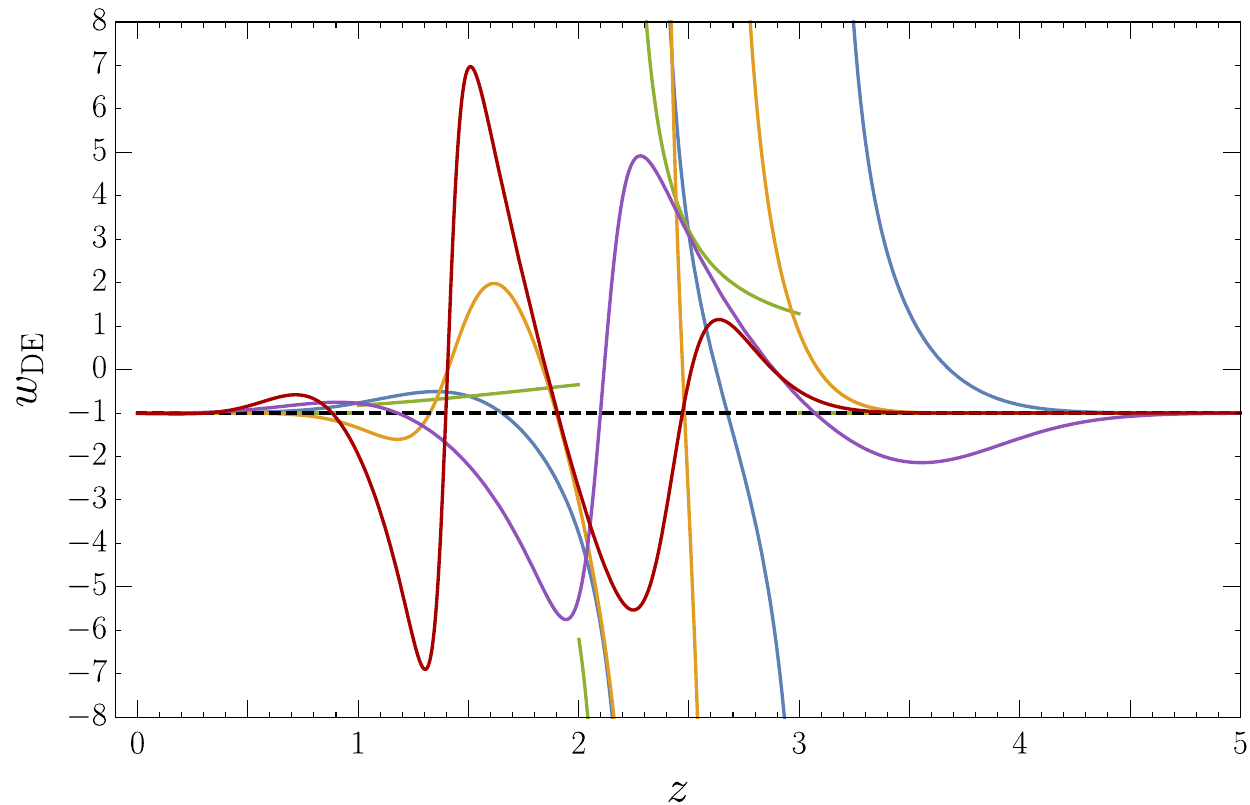}
    \includegraphics[width=0.45\textwidth]{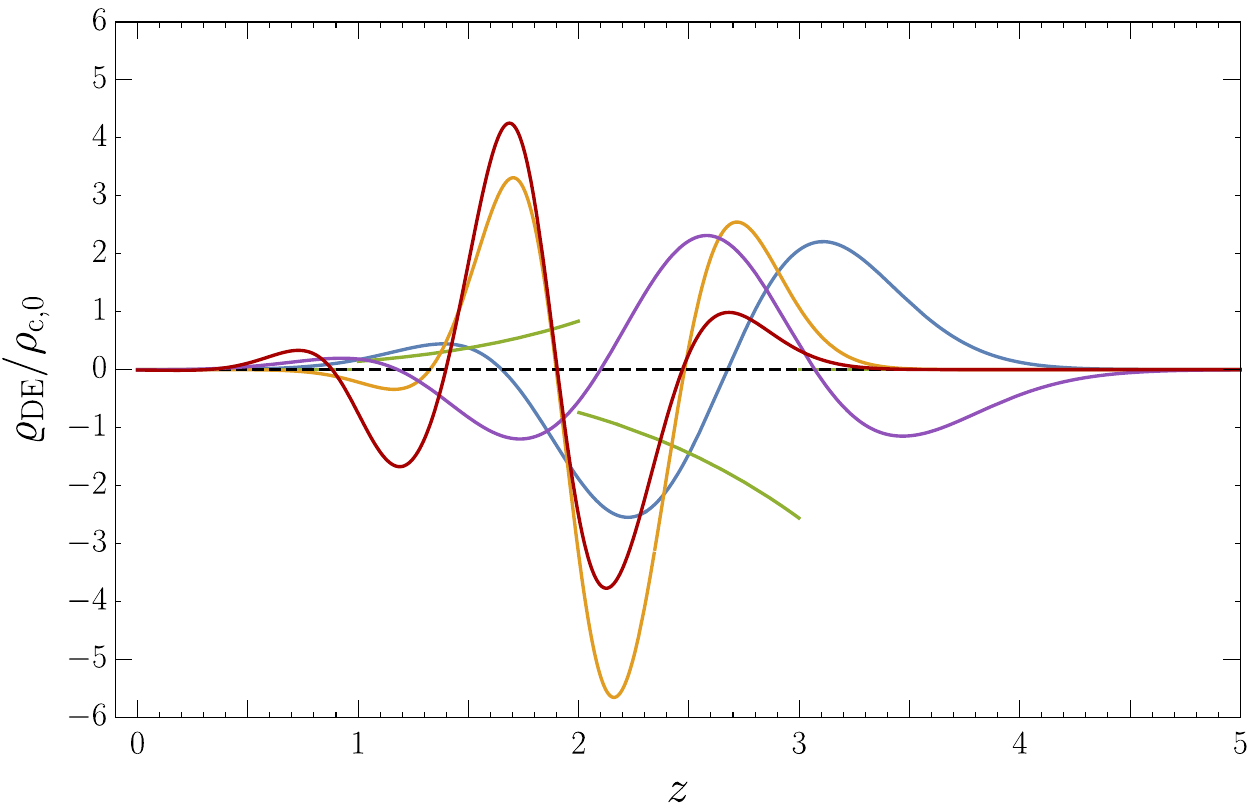}
    \caption{The deviations from the cosmological constant, if the wavelet examples of $\psi(z)$ are attributed to a dynamical DE, i.e., the wiggles in $H(z)$ are produced solely by a dynamical DE. The plots are matched by color to those in \cref{fig:hradius}.}
    \label{fig:dens}
\end{figure*}

\begin{figure}[t!]
    \begin{flushright}
    \includegraphics[width=0.45\textwidth]{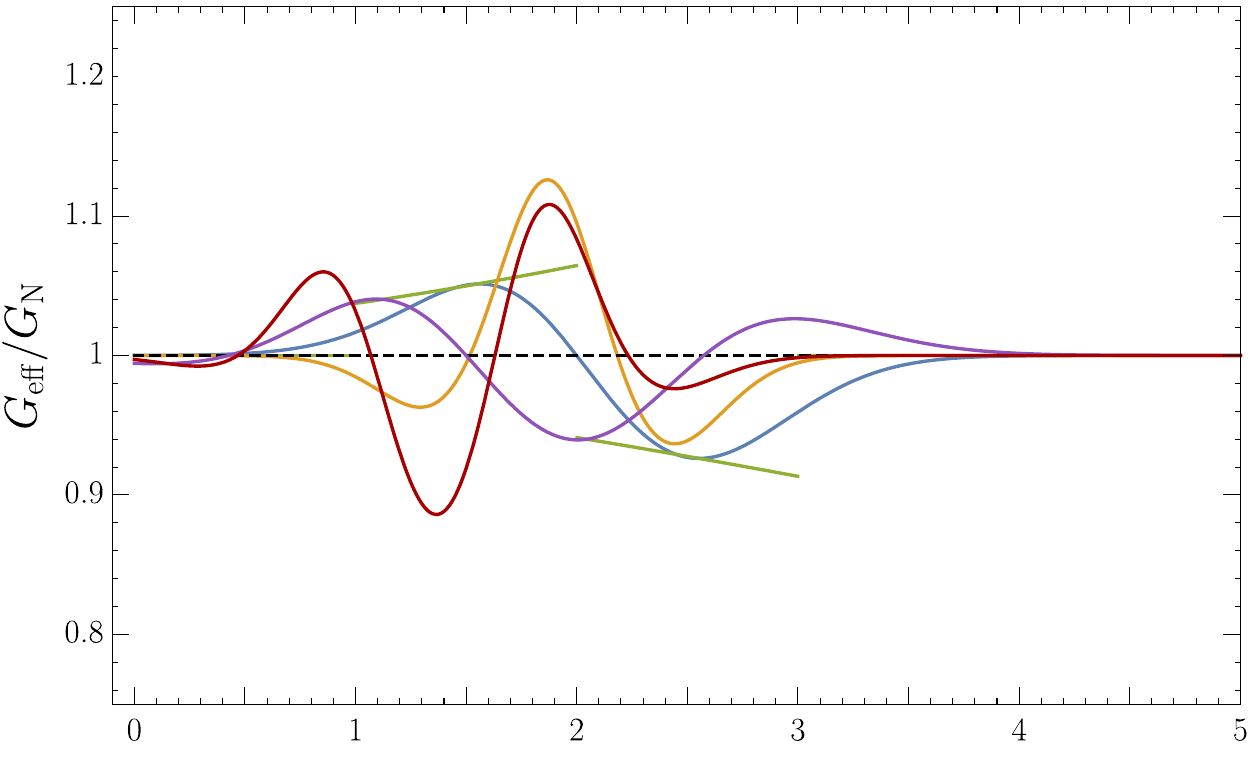}
    \includegraphics[width=0.467\textwidth]{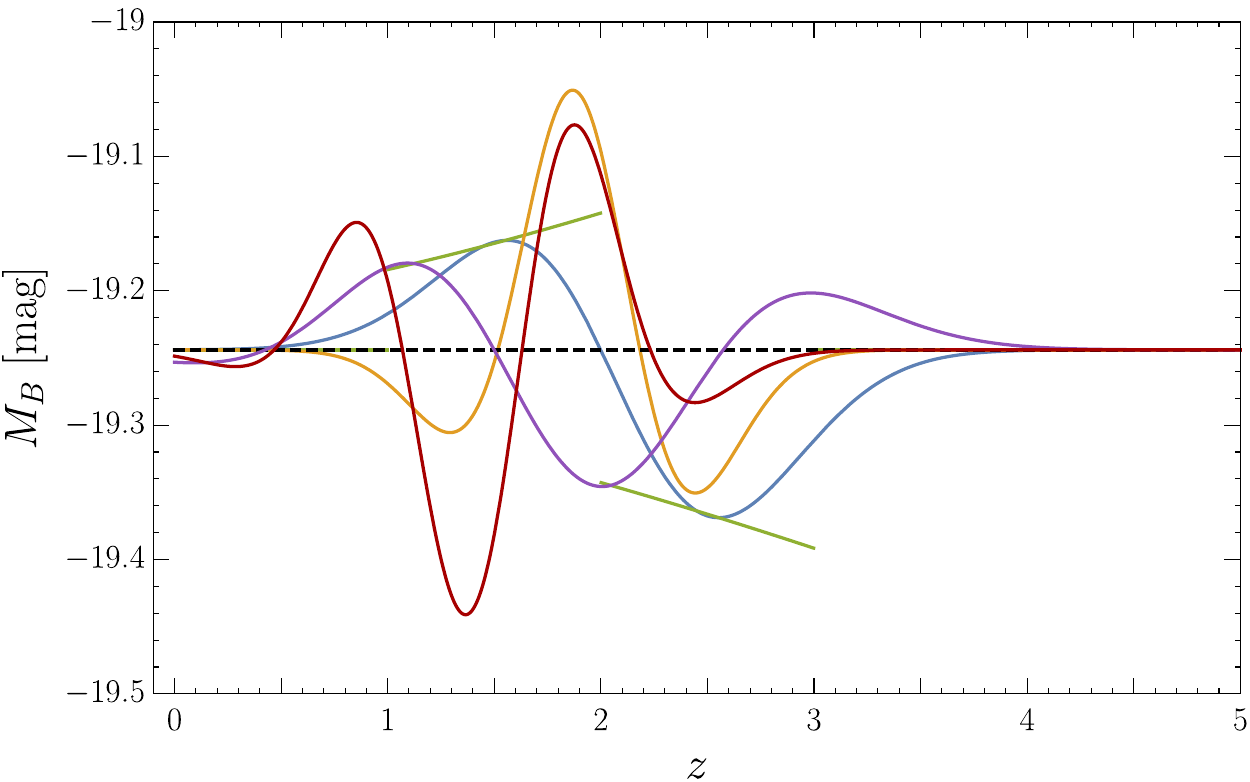}
    \end{flushright}
    \caption{ The deviation from $G_{\rm N}$ if wiggles are produced solely by a varying Newton's ``constant"; we also show the variation in the absolute magnitude $M_B$ of supernovae assuming the unmodified value to be the mean value of the measurement in Ref.~\cite{Camarena:2021jlr}. The variation of $G_{\rm eff}$ is less than $\sim10\%$ at all times, and there is practically no variation for $z\sim0$ and $z\gg0$.
    The plots are matched by color to those in \cref{fig:hradius}.}
    \label{fig:ex}
\end{figure}

These example wavelets, and their corresponding cosmologically relevant functions are plotted in~\cref{fig:hradius,fig:dens,fig:ex,fig:kinematics} for various values of their free parameters; matching colors in different figures indicate the same wavelet with the same choice of parameters. The dashed line corresponds to a vanishing wavelet, i.e., to the reference $\Lambda$CDM model described with $H_{\Lambda\rm CDM}(z)$; for the figures, we neglected radiation for $z<z_*$, and used the mean values of the Planck 2018 TT,TE,EE+lowE+lensing results~\cite{Planck:2018vyg} for $\rho_{\rm m0}/H_0^2$, $r_*$, $\theta_*$, and $z_*$. \cref{fig:hradius} shows the wavelets themselves on the bottom panel and the corresponding Hubble radii on the top panel. \cref{fig:kinematics} shows some cosmological kinematics related to the same wavelets. The top left panel shows the Hubble parameter, $H(z)$, and the top right panel shows the comoving Hubble parameter, viz., the expansion speed $\dot{a}=H(z)/(1+z)$, where $a$ is the scale factor of the spatially flat RW metric, and dot denotes ${\rm d}/{\rm d}t$. The lower right panel shows $D_M^{-1}(z)$ scaled by $c\ln(1+z)$. The data points in these right panels are the local ${H_0=69.8\pm0.8\rm \,km\, s^{-1}\, Mpc^{-1}}$ measurement utilizing the tip of the red giant branch (TRGB)~\cite{Freedman:2019jwv}, and the BAO measurements (see Ref.~\cite{eBOSS:2020yzd} and references therein): BOSS DR12 consensus Galaxy (from $z_{\rm eff}=0.38,\,0.51$), eBOSS DR16 LRG (from $z_{\rm eff}=0.70$), eBOSS DR16 Quasar (from $z_{\rm eff}=1.48$), eBOSS DR16 Ly-$\alpha$-Ly-$\alpha$ (from $z_{\rm eff}=2.33$), and eBOSS DR16 Ly-$\alpha$-quasar (from $z_{\rm eff}=2.33$ but shifted to $z=2.35$ in the figures for visual clarity). The lower left panel shows the derivative of the Hubble function with respect to the cosmic time normalized by $3H^2$, a crossing of the zero would indicate a nonmonotonic behaviour in the Hubble function as suggested in~\cite{Aubourg:2014yra,Akarsu:2019hmw,Akarsu:2021fol}. \cref{fig:dens} shows various plots related to the DE dynamics when the wiggles are attributed to the DE. The top left panel shows the DE densities normalized by the present-day  critical energy density $\rho_{\rm c0}$, the top right panel shows the DE density parameters, $\Omega_{\rm DE}(z)\equiv \rho_{\rm DE}(z)/3H^2(z)$, the lower left panel shows the corresponding DE EoS parameters, and the lower right panel shows the DE inertial mass densities, $\varrho_{\rm DE}(z)\equiv\rho_{\rm DE}(z)+p_{\rm DE}(z)$, normalized by $\rho_{\rm c0}$. Finally, \cref{fig:ex} shows the corresponding results when wiggles in $H(z)$ are attributed to $G_{\rm eff}(z)$. The top panel shows $G_{\rm eff}(z)$ normalized by $G_{\rm N}$, say $G_{\rm eff}(z=0)$, and the lower panel shows the variation of the absolute magnitude $M_B(z)$ for the mean value of the measurement $M_{B,G_{\rm N}}=M_{B}(z=0)=-19.244\pm0.037$ mag inferred in Ref.~\cite{Camarena:2021jlr} (using the Pantheon SnIa data set~\cite{Pan-STARRS1:2017jku} along with Cepheid stars at $z<0.01$ for their calibration).

\subsection{Haar wavelet}
\label{sec:haar}
To begin with, we consider the simplest wavelet, the Haar mother wavelet~\cite{Chui:1992}; namely, $\psi_{\rm h}(z)=1$ for ${0\leq z< 1/2}$, ${\psi_{\rm h}(z)=-1}$ for $1/2\leq z<1$, and zero everywhere else; so that $\Psi_{\rm h}(z_*)=0$.  Shifting and scaling $\psi_{\rm h}$ with three parameters, i.e., defining 
\begin{equation}
   \psi_{\rm H}(z)\equiv \Bar{\alpha} \psi_{\rm h}\qty[\Bar{\beta}\qty(z-\Bar{z}_\dagger)+\frac{1}{2}],
\end{equation}
we can produce discontinuous wiggles on $H(z)$. For a good description of the Ly-$\alpha$ data, we choose ${\Bar{\alpha}=0.00015}{\rm \,km\, s^{-1}\, Mpc^{-1}}$, ${\Bar{z}_\dagger=2}$, and ${\Bar{\beta}=\frac{1}{2}}$ for our example; see the green lines in the figures. It is clear that, for these values of its parameters, $\psi_{\rm H}(z)$ satisfies all the conditions we imposed on $\psi(z)$, the major ones being~\cref{eq:int,eq:prereccond,eq:present}. In the figures, $\delta(z)<0$ for the interval $z\in[2,3)$ leads to a dip in $H(z)$ for this interval that is in excellent agreement with the Ly-$\alpha$ data. This is compensated by $\delta(z)>0$ for the interval $z\in[1,2)$ so that~\cref{eq:int} is satisfied; this region constitutes a bump on $H(z)$. This bump presents itself in other functions such as $\rho_{\rm DE}(z)$ and $\varrho_{\rm DE}(z)$ [or $G_{\rm eff}(z)$], and it is reminiscent of those that are found in nonparametric DE density reconstructions~\cite{Escamilla:2021uoj,Bernardo:2021cxi} from observational data. For our particular example, we see in the figures that the bump results in slight disagreement with the eBOSS DR16 Quasar data at $z_{\rm eff}=1.48$ for both $H(z)$ and $D_{M}(z)$. This can be mitigated by a different choice of parameters or more interestingly by adding more wiggles, for example, by superposing multiple Haar wavelets; however, this superposition would increase the number of free parameters. In the next subsection, we will increase the number of wiggles without increasing the number of free parameters. Note that, the $\Dot{H}(z)/3H^2(z)$ plot of the Haar example appears to never cross the zero line, implying monotonic behaviour for $H(z)$; however, this is not true. The discontinuities of the $H(z)$ function at $z=1,2,3$ result in spikes (Dirac delta distributions) that are not shown in~\cref{fig:kinematics} for the $\dot{H}(z)$ function at these redshifts, resulting in two crossings of the zero line at $z=1,2$ and a nonmonotonic $H(z)$ that increases instantaneously in time at $z=2$. Similar spikes also exist for the $w_{\rm DE}(z)$ and $\varrho_{\rm DE}(z)$ functions if the wiggles are attributed to the DE, but again are not shown in \cref{fig:dens}. Additionally, if the deformations of the Hubble function described by $\delta(z)$ is attributed to the DE density, the $w_{\rm DE}(z)$ has a discontinuity at $z\sim 2.2$ (as suggested in~\cite{Akarsu:2019hmw,Akarsu:2021fol,Akarsu:2022typ}) in addition to the obvious ones at $z={1,2,3}$. This discontinuity (present as a singularity) happens exactly at the redshift in which $\rho_{\rm DE}(z)$ crosses from negative to positive values, and is characteristic of energy densities that have vanishing values in time and not problematic from the point of view of fundamental physics as discussed below~\cref{eq:eos}. Of course, the discontinuities at $z={1,2,3}$ are not very compelling physically, but the Haar wavelet is the simplest example and shows what we should expect from the form of $H(z)$ for a minimal wavelet type deviation of $H(z)^{-1}$ from $H_{\Lambda \rm CDM}(z)^{-1}$. A good alternative to the Haar wavelet can be the Beta wavelet~\cite{betawave} derived from the derivative of the Beta distribution ${P_\beta(z|\gamma,\lambda)\equiv1/B(\gamma,\lambda)z^{\gamma-1}(1-z)^{\lambda-1}}$ where $B(\gamma,\lambda)\equiv\int_0^1 k^{\gamma-1}(1-k)^{\lambda-1}\dd{k}$ is the Euler beta function, $0\leq z\leq1$, and $1\leq\gamma,\,\lambda\leq\infty$. Beta wavelets are in some sense softened Haar wavelets as both have compact support and are unicycle (i.e., they have just one bump and one dip), however, unlike the Haar wavelet, the Beta wavelet is continuous~\cite{betawave}. Thus, to describe more wiggles, one would need to superpose multiple Beta wavelets just like in the case of the Haar wavelet, increasing the number of free parameters. While the Beta wavelets can satisfy \cref{eq:int,eq:prereccond,eq:present} exactly without compromising continuity, they do not have a closed-form expression and are mathematically less tractable; thus, for simplicity, we will proceed with Hermitian wavelets that are also continuous\footnote{One may also wish the wavelet satisfying~\cref{eq:int,eq:prereccond,eq:present} exactly to have the stronger property of being smooth. However, since these conditions require that every derivative of the wavelet vanish outside of the interval $[0,z_*]$, but not inside, such a wavelet cannot be analytic. Nonanalytic smooth functions can be constructed piecewise similarly to splines but the pieces are not necessarily polynomial. These kinds of functions are not compelling for the demonstrative purposes of this paper but may turn out to be useful in observational analyses.} and simpler, and satisfy~\cref{eq:int,eq:prereccond,eq:present} to high precision.

\subsection{Hermitian wavelets}
\label{sec:hermitian}
The discontinuous features of the Haar wavelet can be considered as an approximate description of a rapidly varying smooth function which would be physically more relevant. A simple family of smooth wavelets can be acquired from the derivatives of a Gaussian distribution (cf., the Hermitian wavelets~\cite{hermitianwave}). To do so, we consider the Gaussian distribution defined as follows:
\be
\label{eqn:G0}
{\psi_{\rm G0}(z)=-\frac{\alpha}{2\beta}e^{-\beta(z-z_\dagger)^2}},
\ee
where $\alpha$, $\beta>0$, and $z_\dagger>0$ are the three free parameters that will set, respectively, the amplitude, support, and center of the wiggles. The real part of the $n$th Hermitian wavelet can be obtained from the $n$th derivative of a Gaussian distribution ${\psi_{\text{G}n}(z)\equiv\dv[n]{\psi_{\rm G0}(z)}{z}}$; accordingly, utilizing~\cref{eqn:G0} we obtain
\begin{equation}
\begin{aligned}
\psi_{\rm G1}(z)=&-2\beta(z-z_{\dagger}) \psi_{\rm G0}(z),\\
\psi_{\rm G2}(z)=&4\beta \qty[\beta(z-z_{\dagger})^2-\frac{1}{2}]
\psi_{\rm G0}(z), \\
\psi_{\rm G3}(z)=&-8\beta^2 
\qty[\beta(z-z_{\dagger})^3-\frac{3}{2}(z-z_\dagger)]\psi_{\rm G0}(z), \\
\psi_{\rm G4}(z)=&16\beta^2\qty[\frac{3}{4}+(z-z_{\dagger})^4\beta^2-3\beta(z-z_{\dagger})^2]\psi_{\rm G0}(z),
\end{aligned}
\end{equation}
etc., where only up to fourth derivative are written explicitly. $\psi_{\rm G1}(z)$ and $\psi_{\rm G2}(z)$ are well-known wavelets and the latter is also known as the Ricker (Mexican hat) wavelet. $\psi_{\text{G}n}(z)$ are quasiperiodic functions, i.e., the redshift difference between consecutive peaks (whose amplitudes may differ) of the wave varies. We note that $\psi_{\rm G0}(z)$ itself is responsible for the fast damping of the wavelet function $\psi_{\text{G}n}(z)$ as $z$ moves away from $z_\dagger$ and that $n$th derivative of $\psi_{\rm G0}(z)$ brings an $n$th degree polynomial as a factor to itself, which in turn implies that $n$ stands also for the number of nodes of the $\psi_{\text{G}n}(z)$ function, i.e., the number of times the function crosses zero. These $n$ nodes correspond to $n+1$ wiggles [total of $n+1$ dips and bumps of $\psi(z)$]; the bumps of $\psi(z)$ manifest themselves as dips, and dips of $\psi(z)$ manifest themselves as bumps in $\delta(z)$ and equivalently $H(z)$, cf. Eq.~\eqref{eqn:sdevH}. These manifestations directly translate to wiggles on either $\rho_{\rm DE}(z)$ or $G_{\rm eff}(z)$ depending on which function we attribute them to. The wiggly structure in these functions resemble the wiggles in their respective functions that are acquired from observational analyses utilizing parametric or nonparametric reconstructions~\cite{Escamilla:2021uoj,Bernardo:2021cxi}. Wiggles acquired in observational reconstructions are no surprise even if the dataset does not contain CMB, because wiggles are necessary for $H(z)$ to fit the measurements of the Hubble parameter from the BAO data better than $H_{\Lambda{\rm CDM}}$ without spoiling the success of $\Lambda$CDM in fitting the $D_M(z)$ values measured from the same BAO data (see Fig.~\ref{fig:kinematics}); and the logic we used to show the necessity of bumps still apply when $z_*$ is swapped for the effective redshift of a BAO measurement.

Coincidentally, the first derivative of the Gaussian distribution \eqref{eqn:G0}, i.e., $\psi_{\rm G1}(z)$, can be used to roughly approximate the Haar wavelet smoothly. For $\psi_{\rm G1}(z)$, we pick $\alpha=0.0005{\rm \,km\, s^{-1}\, Mpc^{-1}}$, $z_\dagger=2$, and $\beta=2$, so that the wavelet approximates our previous Haar example. For the rest of the examples, $\psi_{\rm G2}(z)$, $\psi_{\rm G3}(z)$, and $\psi_{\rm G4}(z)$, the values of the parameters are shown on the top panel of~\cref{fig:hradius} and the increased number of wiggles for higher derivatives are clearly seen. Also in~\cref{fig:kinematics}, the top left, top right, and bottom right panels show how increasing the number of wiggles can provide a better description of the BAO data. Unlike the Haar and $\psi_{\rm G1}(z)$ examples, $\psi_{\rm G2}(z)$ and $\psi_{\rm G4}(z)$ examples better describe also the eBOSS DR16 Quasar data at $z_{\rm eff}=1.48$ while retaining better agreement with the Ly-$\alpha$ BAO data at $z_{\rm eff}=2.33$; the $\psi_{\rm G4}(z)$ example even complies with the trend of the Galaxy BAO data (at $z_{\rm eff}=0.38,\,0.51,\,0.70$) $H(z)/(1+z)$ measurements that increase with redshift; this trend is not present in Planck $\Lambda$CDM (i.e., $\Lambda$CDM as constrained by Planck CMB data) even though it is not in strong tension with any of these data points. Still, we emphasize that these wavelets are just illustrative examples and better wavelets can be looked for. Again, attributing the wiggles to the DE, the DE density also wiggles smoothly; however, for the $\psi_{\rm G1}(z)$ and $\psi_{\rm G2}(z)$ examples, two safe/expected singularities are again present in $w_{\rm DE}(z)$ at the redshifts that the DE density vanishes.

Note that \cref{eq:int,eq:prereccond,eq:present} are satisfied exactly only for admissible wavelets with compact support in the redshift interval $[0,z_*]$; thus, unlike the Haar wavelet, $\psi_{\text{G}n}(z)$ does not satisfy~\cref{eq:int,eq:prereccond,eq:present} exactly, but rather approximately\footnote{We emphasize that the Haar and Hermitian wavelets are just convenient examples we used to demonstrate various aspects of the wavelet framework. The previously mentioned Beta wavelets can satisfy these conditions exactly without compromising continuity (at the cost of simplicity due to their lack of closed-form expression). Additionally, working with wavelets generated by higher order derivatives of the Beta distribution, it should be possible to increase the number of wiggles without increasing the number of free parameters, but to our knowledge, there is no established literature on wavelets derived from their higher order derivatives. Another possibility is constructing wiggles out of splines that are piecewise polynomials which can have compact support, but these are likely to suffer from an excessive number of free parameters. Also, a middle ground exists where some of the conditions are satisfied exactly and some approximately. For example, the $n$th Poisson wavelet, viz., $\psi_{{\rm P}n}(z)\equiv\frac{z-n}{n!}z^{n-1}e^{-z}$ for $z\geq0$ and vanishing everywhere else, satisfies \cref{eq:present} exactly but the other two equations approximately for $n>1$.} (yet, beyond a level that cannot be resolved by observation). These three conditions were imposed on $\psi(z)$ through arguments relying on the robustness of certain observations; however, no matter how robust and model independent they are, the uncertainties of the measurements themselves require only that \cref{eq:int,eq:prereccond,eq:present} hold approximately. Reassuringly, for large redshifts, ${\psi_{\text{G}n}(z)H_{\Lambda\rm CDM}(z)\propto z^{n+\frac{3}{2}}e^{-\beta z^2}}$ for matter dominated and $\propto z^{n+2}e^{-\beta z^2}$ for radiation dominated universes; both of these functions rapidly decay by virtue of the exponential term which eventually decays faster than any polynomial growth, ensuring $\delta(z)\to0$ at large redshifts; see \cref{eqn:deltaH}. A similar argument can be made for $\Delta\rho_{\rm DE}(z)\to0$ through \cref{eq:deltarho} at large redshifts. Finally, to demonstrate how successfully the $\psi_{\text{G}n}(z)$ examples approximate the conditions given in \cref{eq:int,eq:prereccond,eq:present}, we examine our $\psi_{\rm G3}(z)$ example as it is the one that violates these conditions most strongly. The values we pick in our $\psi_{\rm G3}(z)$ example correspond to the following quantities related to \cref{eq:int,eq:prereccond,eq:present}: $\psi_{\rm G3}(0)=(41.75\times10^{-6})\,{\rm \,km^{-1}\, s\, Mpc}$ [related to \cref{eq:present}] which can be compared with ${H^{-1}_{\Lambda\rm CDM}(0)=(14.78\times10^{-3})\,{\rm \,km^{-1}\, s\, Mpc}}$ from Planck 2018 resulting in ${\delta(0)\sim3\times10^{-3}}$, $\psi_{\rm G3}(z_*)\sim 10^{-10^6}{\rm \,km^{-1}\, s\, Mpc}$ [related to \cref{eq:prereccond}] which can be compared with ${H^{-1}_{\Lambda\rm CDM}(z_*)=7.3\times10^{-7}{\rm \,km^{-1}\, s\, Mpc}}$ from Planck 2018 resulting in ${\delta(z_*)\sim10^{-10^6}}$, and ${c\times\Psi_{\rm G3}(z_*)=-5.46\,\rm Mpc}$ [related to \cref{eq:int}] which is extremely well within the $1\sigma$ uncertainty of $D_M(z_*)=13872.83\pm 25.31\,\rm Mpc$ measured in Planck 2018.

\section{Conclusion}
It is well known that the comoving angular diameter distance to the last scattering surface, $D_M(z_*)$, is strictly constrained by observations almost model independently. Therefore, in a viable cosmological model, this distance should be the same with the one measured by assuming $\Lambda$CDM, so that consistency with CMB data is ensured at the background level. We have shown mathematically in \cref{sec:statement} that, assuming the prerecombination and present-day universes are well described by $\Lambda$CDM, this is satisfied only if the deviation of any model from $\Lambda$CDM described by the function $\psi(z)=H^{-1}(z)-H_{\Lambda\rm CDM}^{-1}(z)$, which is the deviation from the standard $\Lambda$CDM model's Hubble radius, is an admissible wavelet or is well approximated by an admissible wavelet. In other words, in a viable alternative cosmological model that leaves the prerecombination and present-day universes as they are in the standard cosmological model, the modifications cannot be arbitrary but should satisfy (exactly or approximately at a precision level that can be absorbed within the precision of the available observational data) a Hubble radius function whose deviation from the one in the standard cosmological model is a member of the set of admissible wavelets.

The admissible wavelets describing $\psi(z)$ can be converted to modifications in various cosmological kinematic functions such as the Hubble and comoving Hubble parameters, $H(z)$ and $H(z)/(1+z)$ as shown in \cref{fig:kinematics}, as well as the deceleration and jerk parameters, $q(z)$ and $j(z)$. The wiggly nature of wavelets describing $\psi(z)$ leads to wiggles in these functions, but none of them are necessarily wavelets, moreover, even the ones that arise from the simplest wavelets have nontrivial behaviour that is highly unlikely to be constructed/introduced by hand in the first place. Accordingly, requiring $\psi(z)$ to be an admissible wavelet not only ensures consistency with the CMB at the background level, but also the corresponding wiggles coming on top of the kinematic functions of $\Lambda$CDM can provide us with an improved description of the multitude of BAO data compared to $\Lambda$CDM; as can be seen in \cref{fig:hradius,fig:kinematics}. 
Also, as the wavelets we used as examples show, the number of wiggles in $\psi(z)$, hence also in cosmological kinematics, can be varied and then featured kinematics well fitting the observational data can be achieved without further increasing the number of free parameters; e.g., one may introduce any number of wiggles by taking a sufficient number of derivatives of the Gaussian distribution and still have only three extra free parameters. These nontrivial modifications we have found in the cosmological kinematics can then be attributed to different physical origins. As the first examples that come to mind, we have attributed them either to a dynamical DE, viz., $\rho_{\rm DE}(z)$, in \cref{sec:de_descend} or to a dynamical gravitational coupling strength, viz., $G_{\rm eff}(z)$, in \cref{sec:g_descend} and briefly discussed how these different approaches are, in principle, observationally distinguishable, even though they give rise to the same background kinematics---see \cref{fig:dens,fig:ex} showing what kind of behaviours the example wavelets correspond to in both cases. We demonstrated also that the dynamics of the DE, or the gravitational ``constant'', led by the simplest wavelets, are even more nontrivial compared to the kinematics; for instance, the DE density can change sign in the past, accompanied by singularities in its EoS parameter.

Some phenomenological studies find wiggly structures in various cosmologically relevant functions, and the wavelet framework suggests also being cautious when attributing a physical reality to these wiggles, see, e.g., ~\cite{Colgain:2021pmf,Pogosian:2021mcs,Raveri:2021dbu,Wang:2018fng,Escamilla:2021uoj,Bernardo:2021cxi,Tamayo:2019gqj}. A wiggly structure may be described as consecutive bumps and dips on a function. By using the simplest admissible wavelets employed as examples in~\cref{sec:ex}, we encountered a common pattern that these toy examples, which well-describe the BAO data, present a bump in the Hubble parameter (which can be attributed to a bump in the DE density) at $1.5\lesssim z\lesssim2$ just as was found in various observational reconstructions~\cite{Wang:2018fng,Escamilla:2021uoj,Bernardo:2021cxi}. The existence of bumps is a natural outcome of our findings, because the dips in $H(z)$ required for a better description of the present data, e.g., at $z\sim2.3$ relevant to the Ly-$\alpha$ data, should be compensated by bumps elsewhere so that the comoving angular diameter distance to the last scattering surface remains unaltered. This should raise serious concerns that the bumpy features in the nonparametric $H(z)$ and/or $\rho_{\rm DE}(z)$ reconstructions may be fake in two ways. First, the compensatory bumps could appear at redshifts at which there are no data points to oppose the bumps; it would not be possible to pin down the time location of a bump (or multiple bumps) without new observations. However, most observational analyses reconstruct the cosmological functions up to $z\sim3$ where the most suitable redshift range for a fake bump appears to be at $1.5\lesssim z\lesssim2$, whereas the redshift range devoid of data where these bumps may be present is actually arbitrary and can extend to very high redshifts (e.g., a plateau with a small amplitude over a large redshift range compensating a tight dip at $z\lesssim3$). Second, it may be the case that even the precedent dip that the bump compensates is artificial, e.g., the dip may be caused by overfitting to the data, or the data calling for the dip (e.g., Ly-$\alpha$ BAO) itself may be suffering from systematic errors; in these cases, both the dip and the bump could be fake. It is worth noting here that the wiggles in the DE density are not expected to be representative of an effective field theory, more concretely any minimally coupled scalar model~\cite{Colgain:2021pmf}, and thus it is conceivable that the introduction of theoretical priors should smooth out the wiggles in the DE density~\cite{Pogosian:2021mcs,Raveri:2021dbu}. This may be implying that, if they are real, the origin of the wiggles in $H(z)$ must be sought in modified gravity theories. However, it may also be too hasty to completely ignore the possibility of finding highly wiggly (may be discrete) DE densities; see, for instance, the so-called Everpresent $\Lambda$ model, which suggests the observed $\Lambda$ fluctuates between positive and negative values with a magnitude comparable to the cosmological critical energy density about a vanishing mean, $\braket{\Lambda}=0$, in any epoch of the Universe, in accordance with a long-standing heuristic prediction of the causal set approach to quantum gravity~\cite{Ahmed:2002mj,Zwane:2017xbg,Surya:2019ndm}.

Up until now we have avoided discussing the $H_0$ tension and assumed that any alternative cosmological model would not deviate from $\Lambda$CDM at $z\sim0$ based on the observational argument that $\Lambda$CDM describes local observational data well and is also supported by nonparametric reconstructions. However, this no deviation condition [cf. \cref{eq:present}] is stricter than necessary, because it is essentially the functional form of ${3H^2_{\Lambda\rm CDM}(z)=\rho_{\rm m,0}(1+z)^3+\rho_{\Lambda}}$ that is favored by local data. This suggests that the reference model from which the deviations are defined can be taken to be any model that is compatible with CMB data while agreeing with the functional form of $\Lambda$CDM exactly or approximately in the vicinity of the present-time of the Universe, instead of the exact $\Lambda$CDM model itself. Such models can be compatible with both CMB and local $H_0$ measurements at the same time, see e.g., Refs.~\cite{Akarsu:2019hmw,Akarsu:2021fol,Akarsu:2022typ}. Even the requirement of this functional form can be relaxed and the well-known CPL parametrization and $w$CDM model can be used for the reference model, in which case $\psi(z)$ being an admissible wavelet is not a necessary condition but an analytically compelling case. Even though the functional form of such alternative reference models allows them to simultaneously fit the CMB and $H_0$ measurements, it is possible that strict observational constraints from BAO data prevent these models from occupying the part of their parameter space required for this simultaneous agreement. If these models are taken to be the reference model, the $H_0$ tension may also be resolved within our wavelet framework; more importantly, if the observational success of these models were held back by the BAO data, the use of wavelets may resurrect them by letting them fit the BAO data without compromising their successful description of the CMB and $H_0$ observations.

In our discussions we allowed wavelets to have quite a bit of freedom, apart from requiring them to be admissible and vanish outside of the interval $z=[0,z_*]$, see \cref{eq:int,eq:prereccond,eq:present}. However, it can also be very useful to focus on various subsets of these wavelets. Namely, using arguments based on the history of the expansion of the Universe and/or fundamental physics (also, these two can be related in a certain way through the putative theory of gravity), we can impose more conditions on them, and thereby narrow down the extent of the family of cosmological models satisfying our conditions. For example, as we have already discussed to some extent, with regard to the kinematics of the Universe, one may demand an ever expanding universe [$H(z)>0$] and/or a monotonically decreasing Hubble parameter [$\dot{H}(z)<0$] from beginning to the present, or, with regards to dynamics of the DE (supposing that GR is valid and the deviations are attributed to a dynamical DE fluid), one may demand a non-negative DE density [$\rho_{\rm DE}(z)\geq0$] at all times, or a non-negative DE inertial mass density corresponding to the null energy condition ($\varrho_{\rm DE}(z)\geq0$) at all times, or at least be cautious so that no instability problems are encountered. Indeed, DE fluids that lead to our example admissible wavelets, seem to easily violate the conventional energy conditions; namely, the EoS parameter crosses below minus unity and/or plus unity and even exhibits  poles in some cases, moreover, these behaviours correspond to a DE inertial mass density that crosses below zero, and even a DE density that crosses below zero for the cases whose EoS parameter exhibits poles. Such violations are generally known to indicate possible instability issues in the DE fluid. One way out in this case, as we mentioned earlier, would be the possibility of deriving such dark energies from modified gravity theories as effective sources without causing some other instability problems. Employing the parameterized post-Friedmann~\cite{Hu:2008zd, Fang:2008sn} approach may also provide us with another way out, namely, the parameterized post-Friedmann approach discussed in~\cite{Hu:2008zd, Fang:2008sn} may be used to placate the violent behaviors of the DE source, particularly to solve the instability issues related to the DE EoS parameter or make them less severe by pulling it towards the safer interval $[-1,1]$. This approach that replaces the condition of DE pressure perturbation with a smooth transition scale will help us understand the momentum density of the DE and other components on the large scale structure. We leave the advantages of considering such reconstruction methods in relevant to the family of the DE models introduced in this paper for future consideration.

To conclude, the wavelet framework presented in this paper seems to have the potential to be a good guide to find new cosmological models, alternative to the base $\Lambda$CDM model, that are consistent with the observational data and to analyze existing ones, but further observational and theoretical studies are required to uncover the full scope of the implications and applications of this framework.

\begin{acknowledgments}
 The authors thank to Bum-Hoon Lee and Kazuya Koyama for useful insights and discussions. \"{O}.A. acknowledges the support by the Turkish Academy of Sciences in the scheme of the Outstanding Young Scientist Award  (T\"{U}BA-GEB\.{I}P). E.\'O.C. was supported by the National Research Foundation of Korea grant funded by the Korea government (MSIT) (NRF-2020R1A2C1102899). E.\"{O}.~acknowledges the support by The Scientific and Technological Research Council of Turkey (T\"{U}B\.{I}TAK) in the scheme of 2211/A National PhD Scholarship Program.  S.T. and L.Y. were supported by Basic Science Research Program through the National Research Foundation of Korea (NRF) funded by the Ministry of Education through the Center for Quantum Spacetime (CQUeST) of Sogang University (NRF-2020R1A6A1A03047877). L.Y. also thanks the support from YST project in APCTP.
 \end{acknowledgments}

\end{document}